\documentclass[onecolumn,amsmath,amssymb,amsfonts,showkeys,preprintnumbers,a4paper,nofootinbib,superscriptaddress]{revtex4}

\usepackage{graphicx}
\usepackage{color}
\definecolor{orange}{RGB}{255,127,0}
\usepackage{bm}
\usepackage{epstopdf}
\usepackage{url}

\def\gtsima{$\; \buildrel > \over \sim \;$}
\def\ltsima{$\; \buildrel < \over \sim \;$}
\def\gsim{\lower.5ex\hbox{\gtsima}}
\def\lsim{\lower.5ex\hbox{\ltsima}}

\newcommand{\GeV}{{\rm GeV}}

\newcommand{\kpc}{{\rm kpc}}

\newcommand{\m}{{\rm m}}
\newcommand{\cm}{{\rm cm}}
\newcommand{\km}{{\rm km}}

\newcommand{\s}{{\rm s}}

\newcommand{\sr}{{\rm sr}}

\newcommand{	\sv}{\langle\sigma v\rangle}

\newcommand{\ds}{{\sffamily DarkSUSY}}
\newcommand{\dragon}{{\sffamily DRAGON}}
\newcommand{\pythia}{{\sffamily PYTHIA}}
\newcommand{\galprop}{{\sffamily GALPROP}}

\begin{document}

\title{Antiprotons from dark matter annihilation in the Galaxy: astrophysical uncertainties}

\author{Carmelo Evoli}
\email{carmelo.evoli@desy.de}
\affiliation{National Astronomical Observatories, Chinese Academy of Sciences, 20A Datun Road, Beijing 100012, P.R. China} 
\affiliation{Universit\"at Hamburg, Luruper Chaussee, 149 22761 Hamburg, Germany}
\author{Ilias Cholis}
\email{ilias.cholis@sissa.it}
\affiliation{SISSA and INFN, Sezione di Trieste, Via Bonomea, 265, 34136 Trieste, Italy}
\author{Dario Grasso}
\email{dario.grasso@pi.infn.it}
\affiliation{INFN Sezione di Pisa, Largo Bruno Pontecorvo 3, 56127 Pisa, Italy}
\affiliation{Dipartimento di Fisica, Universit\'a di Siena, Via Roma 56, I-56100 Siena, Italy}
\author{Luca Maccione}
\email{luca.maccione@lmu.de}
\affiliation{DESY, Theory Group, Notkestra{\ss}e 85, D-22607 Hamburg, Germany}
\affiliation{Arnold Sommerfeld Center, Ludwig-Maximilians-Universit\"at, Theresienstra{\ss}e 37, 80333 M\"unchen, Germany} 
\affiliation{Max-Planck-Institut f\"ur Physik, F\"ohringer Ring 6, 80805 M\"unchen, Germany} 
\author{Piero Ullio}
\email{ullio@sissa.it}
\affiliation{SISSA and  INFN, Sezione di Trieste, Via Bonomea, 265, 34136 Trieste, Italy}

\date{\today}

\begin{abstract}
{The latest years have seen steady progresses in WIMP dark matter (DM) searches, with hints of possible signals suggested both in direct and indirect detection. Antiprotons play a key role in this context, since WIMP annihilations can be a copious source of antiprotons, and the antiproton flux from conventional astrophysical sources is predicted with fair accuracy and matches the measured cosmic ray (CR) spectrum very well. Using the publicly available numerical \dragon\ code, we reconsider antiprotons as a tool to set constraints on DM models; we compare against the most up-to-date $\bar{p}$ measurements, taking also into account the latest spectral information on the $p$ and $He$ CR fluxes. In particular, we probe carefully the uncertainties associated to both standard astrophysical and DM originated antiprotons, by using a variety of distinctively different assumptions for the propagation of CRs and for the DM distribution in the Galaxy. We find that the impact of the astrophysical uncertainties on constraining the DM properties of a wide class of annihilating DM models can be much stronger, up to a factor of $\sim50$, than the one due to uncertainties on the DM distribution ($\sim 2-6$). Remarkably, even reducing the uncertainties on the propagation parameters derived by local observables, non-local effects can change our predictions for the constraints even by 50\%. Nevertheless, current $\bar{p}$ data can place tight constraints on DM models, excluding some of those suggested in connection with indirect and direct searches. Finally we discuss the impact of upcoming CR spectral data from the AMS-02 instrument on DM model constraints.}
\end{abstract}

\preprint{DESY 11-135}

\maketitle

\section[]{Introduction}

The identification of the nature of dark matter (DM) in the Universe remains an unsolved problem. Assuming that DM is made of elementary particles, there is unfortunately very scarce information on their properties one can deduce from the very rich observational evidence accumulated from cosmological and astrophysical measurements, on scales ranging from the size of the visible Universe down to subgalactic environments (for a recent review on the DM problem, see, e.g.,~\cite{Bertone:2010zz}). 
While one can exclude that DM is electrically charged, baryonic or hot, hence precluding the possibility that the standard model (SM) of particle physics embeds a DM candidate, there are very loose constraints one can derive on the mass scale of DM particles and their interaction strength with SM states, the two key elements to address the DM detection puzzle. The main guideline has then been to focus on classes of DM candidates which are motivated by a natural mechanism for their generation in the early Universe; in this respect, weakly interactive massive particles (WIMPs) are surely among the leading DM candidates.
As a rule of thumb, a particle with mass in the range between a few GeV and a few TeV has a thermal relic density which is naturally at the level of the measured cosmological density if its coupling to the SM hot plasma is of weak interaction type. The relic abundance scales approximately with the inverse of the thermally averaged pair annihilation cross section into lighter SM particles, and it typically takes the correct value when this is about  $3\times 10^{-26}\;{\rm cm}^3\,{\rm s}^{-1}$. 

The density of WIMPs in today's halos is much smaller than in the early Universe. However there is still a (small but finite) probability for WIMPs to annihilate in pairs and give rise to detectable SM yields. Such indirect DM detection has received a lot of attention in the recent years in connection with the wealth of new data that have become available, especially with the new generation of cosmic- and gamma-ray detectors. Most notably,  there have been a few cases in which possible discrepancies between data and expectations from standard astrophysical components have led to speculations that an extra component due to DM may have been detected: e.g., PAMELA has detected a rise in the positron fraction~\cite{Adriani:2008zr} at energies above about 10~GeV, a result which has been very recently confirmed by the preliminary analysis by the Fermi collaboration \cite{fermieplus}. Also, an excess of gamma-rays has been suggested in an analysis of the Fermi-LAT data in the Galactic center region~\cite{Hooper:2010mq}. Cross correlations with indirect detection channels of a possible detection of WIMP scatterings in the direct detection experiments DAMA-LIBRA~\cite{Bernabei:2008yi} and CoGeNT~\cite{Aalseth:2010vx,Aalseth:2011wp} have also been studied. 

Within such a rich indirect detection program, the role of antiproton measurements has been and remains a major one. There are several aspects why this is the case. First of all, in a ``democratic" WIMP model, namely a scenario in which hadron production in WIMP pair annihilation is not forbidden either by kinematics or some symmetry enforcing WIMPs to be coupled with leptons only, the ratio between DM signal and background from standard astrophysical sources is usually much larger in the antiproton channel with respect to all other indirect detection methods. A second aspect regards the fact that the theoretical prediction for the background component is fairly under control: the production of secondary antiprotons from the interaction of primary cosmic rays (CRs) with the interstellar medium and, subsequently, their propagation in the Galaxy have to be modeled in close analogy to secondary versus primary CR nuclei, such as boron versus carbon. Once a given phenomenological model is tuned to reproduce the latter, the spread in predictions for the antiproton flux is modest. This feature, which has already been discussed, e.g., in Refs.~\cite{Donato01,DiBernardo10}, will be illustrated in further details in this analysis considering a set of radically different physical propagation setups. 

A further aspect making, in principle, the antiproton channel appealing for indirect DM detection, is the fact that background and signal should show readily distinguishable spectral features: By kinematics, the secondary antiproton source function is sharply suppressed at small energies, making the background flux peak at a couple of GeV in kinetic energy; at higher energies the flux settles on a given spectral index, as mainly determined by the spectral index of the primaries and by the dependence on rigidity of the spatial diffusion coefficient. On the other hand, the production of low energy antiprotons is not inhibited in DM annihilations, as well as the DM source function cannot be characterized by an injection power law, but rather as a cascade from a single energy scale, the mass of the annihilating nonrelativistic WIMPs. This will result in a signal with a very broad shape spectrum.~\footnote{The effect of inelastic but non-annihilating scattering of antiprotons with the gas in the interstellar medium, with the production of the so-called {\em tertiary} antiproton component, tends actually to flatten both the astrophysical and DM spectra, thereby broadening the peaked shape of the astrophysical antiproton spectrum and making this argument somewhat looser. However, the qualitative expectations from this shape argument are preserved, as it will be clear in the following.}

The balloon campaigns by the BESS detector~\cite{Matsunaga:1998he,Maeno:2000qx,Asaoka:2001fv} and, even more, the recent measurements by the PAMELA satellite~\cite{Adriani:2010rc} have provided fairly good-precision antiproton data at energies up to about 180~GeV.
A further improvement is soon expected from the AMS-02 observatory \cite{AMS02,ams02proc} on the International Space Station. 
The currently available data indicate quite clearly that the bulk of the local antiproton flux is due to secondaries: there is a close match with the spectral features outlined above for this component, and the normalization of the flux is in good agreement with the predictions within standard CR models fitting secondary and primary CR nuclei. Already at present, the data are very powerful to set constraints on WIMP models, while it is expected that the quality of the data which will be available in the near future will allow to search for slight spectral distortions to be eventually associated to a DM component. It is then timely to reconsider the computation of the antiproton DM signal, discussing in detail the uncertainties involved. Proposed as a signal about 3 decades ago~\cite{Silk:1984zy,Stecker:1985jc}, the DM induced antiproton flux has been computed with different level of sophistication. In the first works the antiproton propagation was treated within the leaky box model, later more realistic two-dimensional diffusion models were implemented (early works include, e.g.,~\cite{Bottino:1998tw,Bergstrom:1999jc}). Under a set of simplifying assumptions (diffusion coefficient and convective winds taken as spatially constant, energy losses and reacceleration effects confined in an infinitely thin disc with constant gas density) the diffusion equation admits a semianalytic solution in the two-dimensional model; this solution is very useful to study systematically the very large parameter space of the model. For a more realistic description of the Galaxy however one needs to implement numerical solutions to the propagation equation, as done, e.g., in \galprop\ or \dragon.

While the issue of uncertainties on the antiproton DM signals has been studied in some details within semianalytic models (see, e.g.,~\cite{Donato01,Donato:2003,Bringmann:2006im,Donato:2008jk}), we present here an extensive study performed with the fully numerical approach. The approach we follow is to introduce a set of rather diverse (and in some aspects extreme) scenarios for the propagation of CRs, setting their properties by fixing some of the parameters in the model, such as the vertical scale for the diffusion coefficient, or its scaling in rigidity, or the strength of the convective winds. In each scenario, using a multidimensional minimization procedure, the additional parameters of the model are fitted against data on the local proton flux (including the recent data from PAMELA) and the boron to carbon ratio. A prediction is then obtained for the background antiproton flux, finding that all models reproduce the data fairly well. 

However, the propagation models that we have introduced have rather diverse properties on a global scale. Therefore, given that the source distribution from DM originated CRs is significantly different from that of more conventional CR sources (for which we fit the propagation properties to the CR data), the impact of the model on the local DM-induced flux can be dramatic, hence introducing rather large uncertainties in their predictions, as we will discuss in this work. While this was already shown with semianalytical models \citep[see e.g.][]{Barrau:2005}, the numerical approach allows to quantify the relative uncertainties in a more general framework.

Moreover, even larger uncertainties can be introduced (and it will be discussed here) when considering nonstandard propagation models, in which some physical processes (e.g.,~diffusion, or convection) do not happen uniformly in the galactic plane, but depend on position.

In this paper we do not consider astrophysical uncertainties which may arise from secondary antiproton production in the SNR surroundings as pointed out in \cite{Blasi:2009} and (like TeV scale DM) could cause a hardening of the antiproton spectrum above 100 GeV.

This paper is organized as follows: in Sec.~\ref{sec:dm} we introduce the DM scenarios we wish to address. In Sec.~\ref{sec:cr_models} we briefly introduce the CR propagation models and the tools we use to solve numerically the CR propagation equation, namely the \dragon\ code~\cite{Evoli:2008dv}. We then define a range of propagation frameworks and their impact on the antiproton flux. In Sec.~\ref{sec:local} we discuss in detail the issue of locality in the secondary and DM-induced source functions with respect to the locally measured antiproton flux; This gives a guideline for a more exotic propagation model that one could consider to maximize the impact on the DM component, as discussed in Sec.~\ref{sec:exotic}. In Sec.~\ref{sec:constraints} we discuss constraints on our selected models within the CR propagation models introduced, while in Sec.~\ref{sec:discuss} we compare with previous results and discuss future perspectives. Section~\ref{sec:concl} is devoted to our final comments and conclusions.

\section[]{Dark matter models}
\label{sec:dm}

There are numerous beyond SM scenarios embedding a WIMP DM candidate. Rather than studying general classes of models over exceedingly large parameter spaces, we chose here to focus on three sample cases which have been recently investigated in connection to hints of DM signals in other detection channels, but potentially giving a sizable antiproton flux as well. These sample cases are also representative of three different WIMP mass regimes, ranging from fairly light models to multi-TeV DM, and are thus sensitive to different parts of the measured antiproton spectrum. 
Since the different assumptions on the galactic CR propagation model influence differently low- or high-energy antiprotons, these three mass ranges are useful to illustrate the dependence of the DM signal on propagation.

\subsection{Nonthermal Wino dark matter}\label{sec:wino}

As a first test case, we consider a pure Wino within the minimal supersymmetric extension to the standard model (MSSM). The Wino is a spin~1/2 Majorana fermion, superpartner of the neutral
SU(2)$_{\rm L}$ gauge boson and one of the four interaction eigenstates whose superposition give rise to  the four neutralino mass eigenstates in the MSSM; we will consider it in the limit when the Wino mass parameter, usually indicated as $M_2$, is much lighter than the other supersymmetry (SUSY) mass parameters, so that interaction and mass eigenstates coincide and the Wino is the lightest SUSY particle (LSP) and, in a R-parity conserving SUSY model, stable. Examples of theories which predict or can embed a low-energy spectrum with a Wino LSP are, e.g., the anomaly mediated SUSY breaking scenario~\cite{Randall:1998uk} and the G2-MSSM~\cite{Acharya:2008zi}.
If kinematically allowed, the Wino pair annihilation is dominated by the $W$~boson final state, driven by the exchange in the t- and u-channel of a Wino-like chargino which, in the pure Wino limit, has also a mass equal to $M_2$, up to a very small mass splitting induced by radiative corrections. Neglecting this small correction, the tree-level cross section for  $\tilde{W}^0\tilde{W}^0 \rightarrow W^+W^-$ in the nonrelativistic limit is given by (see, e.g.,~\cite{Moroi:1999zb}):
\begin{equation}
\sv_{v \rightarrow 0} = \frac{g_2^4}{2\pi} \frac{1}{m_\chi^2}\frac{(1-m_W^2/m_\chi^2)^{3/2}}{(2-m_W^2/m_\chi^2)^{2}},
\label{eq:svwino}  
\end{equation}
where $m_\chi=M_2$ is the Wino mass, $m_W$ the mass of the $W$~boson and $g_2$  the gauge coupling constant of SU(2)$_{\rm L}$.  We will focus on cases with $m_\chi$ in the few hundred GeV range; for such masses, $\sv$ is much larger than the nominal value of about $3\times 10^{-26}\;{\rm cm}^3\,{\rm s}^{-1}$ for thermal relic WIMPs (actually, in this example, this simplified estimate does not hold since chargino coannihilation effects are important, see, e.g.,~\cite{Edsjo:1997bg}; Sommerfeld enhancements, namely long-range effects mediated by SU(2)$_{\rm L}$ bosons, are instead relevant only for much heavier Winos, see, e.g.,~\cite{Hisano:2002fk,Hryczuk:2010zi}). Although the thermal relic component is small, this could still be a viable DM model if Winos are generated nonthermally in the out-of-equilibrium decay of heavy fields, like gravitinos or weakly coupled moduli, see, e.g.,~\cite{Moroi:1999zb,Giudice:2000ex,Lin:2000qq,Acharya:2008bk,Arcadi:2011ev}. In this case the relic density depends on the induced reheating temperature and, possibly, on the branching ratio of the decay into Winos, two quantities that are in turn defined by sectors of the theory we did not specify. We will simply assume that they can be adjusted in such way that any Wino of given mass can be regarded as a good DM candidate. 
Results will also be discussed in the more general scenario in which $m_\chi$ and $\sv$ are assumed as free parameters, but still restricting to the case of $W$~boson as dominant annihilation channel.

The recent interest in this model has been stimulated, besides its peculiar signatures at the LHC, by the claim~\cite{Grajek:2008pg,Kane:2009if} that a Wino with mass of about 200~GeV can explain the rise detected by PAMELA in the positron fraction~\cite{Adriani:2008zr}. This interpretation is controversial since the positron excess that it can indeed induce comes together with a rather copious antiproton yield.
It has been shown that under ``standard" assumptions for cosmic-ray propagation and for the dark matter distribution in the Galaxy, the correlation between the leptonic and hadronic yield of this channel implies that the interpretation of the PAMELA positron data in terms of WIMP annihilating into  $W^+W^-$ is excluded for WIMP masses lighter than a few TeV by the nonobservation of an antiproton excess by PAMELA and in previous antiproton measurements, see, e.g.,~\cite{Cirelli:2008pk,Donato:2008jk,Cholis:2010xb}. 
In \cite{Kane:2009if} three main arguments are given to disregard the antiproton bound: {\sl (i)} Since the positrons from Wino annihilation have, on average, higher energies compared to antiprotons, it should be possible to find some nonstandard energy-dependent propagation setup suppressing the DM-induced antiproton flux, while not affecting the positron signal; {\sl (ii)} The excess in the antiproton flux may stem from a gross overestimation of the secondary antiproton component, while it should be plausible to introduce a model in which the secondary component is subdominant with respect to the DM component, with the latter accounting for the bulk of locally measured antiproton flux; {\sl (iii)} Assuming that the main contribution to the antimatter signals comes from annihilations in dense DM substructures, it should be feasible to find a set of DM point-source configurations for which positrons are favored compared to antiprotons, in connection to the fact that propagation introduces both a scaling with distance and a distortion of the energy spectrum that are different for the two channels. 
We will not reconsider this last issue: it has been shown with semianalytic models, both in the limit of static sources~\cite{Lavalle:1900wn} as well as including proper motion effects~\cite{Regis:2009qt}, that such discreteness effects tends to enhance more the high-energy antiproton flux than the high-energy positron component. 
The second argument will be confuted in the present analysis. As to the first argument, we will show that if one sticks to standard propagation models the antiproton flux is not suppressed enough, even in the most favorable scenario, to allow the DM scenario envisaged in \cite{Kane:2009if}, which can instead be made viable only resorting to nonstandard propagation models (see Sec.~\ref{sec:exotic}).

\subsection{The very heavy WIMP scenario}

Still motivated by the PAMELA positron excess, and possibly in connection with the local all-electron (namely electrons plus positrons) flux measured by Fermi~\cite{Abdo:2009zk} and HESS~\cite{Aharonian:2009ah} and showing  a $E^{-3}$ spectrum hardening at about 100~GeV - 1~TeV, several analyses have considered the possibility of very heavy dark matter WIMPs, with masses up to several TeV and very large pair annihilation cross section, see, e.g.,~\cite{Cirelli:2008pk,Bergstrom:2009fa}. 
The results of such studies are that, to account for the electron/positron component without violating the antiproton bounds, dark matter needs to be leptophilic, i.e., the final products of the annihilation being dominantly leptons, most likely a combination of $e^+ e^-$ and  $\mu^+ \mu^-$. More in general, heavy WIMP models with large annihilation cross section into quark final states are always rather efficiently constrained by antiproton data, since the hadronization of high energy quarks produces a lot of softer antiprotons, i.e., in the energy range covered by PAMELA which extends up to 180~GeV; as mentioned above, final states with weak gauge bosons have also a rich antiproton yield, but the peak in these spectra is shifted to higher energies, so that they may have  escaped detection if the WIMP mass is above a few TeV (the corresponding electron/positron yields fail however to reproduce the spectral features found by Fermi and HESS). 

In most analyses in the literature the antiproton yield from WIMP annihilations has been modeled via Monte Carlo generators like \pythia\ ; it has been recently pointed out~\cite{Ciafaloni:2010ti} that such result is not accurate for very heavy WIMPs because these generators do not include the radiative emission of soft electroweak gauge bosons. This is, in particular, important for the antiproton flux in the case of $W$~boson final states, as well as for leptophilic cases (which have zero antiproton yield in the approximation of two-particle final state).

We will discuss the heavy WIMP regime considering a general framework in which a DM candidate is specified by its mass, the dominant annihilation channel and the pair annihilation rate in the nonrelativistic limit. We will consider a few sample final states, like the antiproton dark matter yield as computed in Ref.~\cite{Ciafaloni:2010ti} including EW corrections, and set upper limits in the mass--cross section plane in the different propagation scenarios.  

\subsection{Light WIMPs with sizable quark couplings}

There have been steady progresses in the field of direct detection in latest years. Most recently, the main focus has been on DM candidates with mass around 10~GeV, with two collaborations having published results compatible with a positive signal: DAMA and DAMA/LIBRA~\cite{Bernabei:2008yi} detected an annual modulation in the total event rate consistent with the effect expected from WIMP elastic scatterings; CoGeNT has just confirmed~\cite{Aalseth:2011wp} the detection of a low-energy exponential tail in their count rate consistent with the shape predicted for the signal from a light WIMP, as already found in a previous data release~\cite{Aalseth:2010vx}, showing in addition a 2.8~$\sigma$ indication in favor of an annual modulation effect. In contrast, CDMS~\cite{Ahmed:2009zw} (see also the recent reanalysis of early data taken at a shallow site~\cite{Akerib:2010pv}), Xenon10~\cite{Angle:2009xb} and, most recently Xenon100~\cite{Aprile:2011hi} have not found any evidence for DM and seem to disfavor the same region of the parameter space.
Taking all data sets at face value, indeed it appears not possible to reconcile them within a single theoretical model (for recent analyses on this point see, e.g.,~\cite{Schwetz:2011xm,Farina:2011pw,Fox:2011px}), indicating there is some missing piece in the puzzle (or some problem with one or more of the datasets or their DM interpretations).
Still, it is interesting to cross correlate with other DM detection techniques.

Under the hypothesis of spin-independent (SI) WIMP-nucleon elastic interactions, a signal at the level of DAMA or CoGeNT requires a fairly large scattering cross section, and hence a sizable coupling between WIMPs and quarks; using crossing symmetry arguments, one expects equally large values of the WIMP pair annihilation cross section into quarks, and hence of the antiproton yields. Considering neutralinos as light as few GeV within a MSSM without the grand unified theory unification condition on gaugino masses, Refs.~\cite{Bottino:2004qi,Bottino:2005xy} noticed a tight connection between a large direct detection signal, at the level of the DAMA modulation effect, and a large antiproton signal, testable in the current generation of cosmic-ray experiments. 
In this specific case, actually, the correlation is to some extent accidental, since it is not the same interaction vertex entering scattering and annihilation and, moreover, the pair annihilation cross section of nonrelativistic Majorana fermions (such as neutralinos) into Dirac fermions is helicity suppressed, i.e., it scales with the square of mass of the final state fermion.

To address the impact of propagation parameters on the antiproton signal from light WIMPs, it is sufficient to consider the simpler framework in which the WIMP-quark interaction is introduced at an effective level integrating out some heavy degrees of freedom, and with the crossing symmetry manifestly implemented. We refer to a case which looks particularly contrived, a real scalar particle $\phi$ with contact interaction contributing to the SI cross section and the annihilation through the quark bilinear $\bar{q}q$~\cite{Bandyopadhyay:2010cc,Andreas:2010dz,He:2010nt,Barger:2010yn,Barger:2010mc}; using the 
same notation as in Ref.~\cite{Keung:2010tu}, this operator is written as
\begin{equation}
{\cal O}_{s} = c_q\, \frac{2 m_\phi}{\Lambda^2} \phi^2 \,\bar{q}q \ .  
\end{equation}  
With this normalization, the SI WIMP-nucleon cross section, usually cast in the form:
\begin{equation}
\sigma_{n,p} = \frac{4}{\pi} \mu_{n,p}^2 f_{n,p}^2 \ , 
\end{equation}
where $\mu_{n,p}$ is the reduced mass of the WIMP-nucleon system (the index $n$ stands for a neutron, the index $p$ for a proton), the effective couplings $f_{n,p}$ are
\begin{equation}
f_{n,p} = \sum_q \frac{c_q}{\Lambda^2} \frac{m_{n,p}}{m_q} f^{(n,p)}_{Tq} \ , 
\end{equation}
where the sum runs over all quarks and the nucleon quark fractions $f^{(n,p)}_{Tq}$ will be assumed according to their mean values in Ref.~\cite{Gasser:1991}. Correspondingly, the pair annihilation cross sections in the nonrelativistic $v \to 0$ limit is given instead by:
\begin{equation}
\sv_{v \rightarrow 0} =\frac{12\, m_\phi^2}{\pi}
\sum_q \left(\frac{c_q}{\Lambda^2}\right)^2 \left(1-\frac{m_q^2}{m_{\phi}^2}\right)^{3/2}\,,
\end{equation}
with the sum running over all quarks lighter than $\phi$. One may consider two extremes: The couplings $c_q$ can be assumed to be universal; in this case the relation between annihilation and scattering cross section WIMP-proton is
\begin{equation}
\sv_{v \rightarrow 0} = \sigma_p \cdot 3 \frac{(m_\phi+m_p)^2}{m_p^2} \frac{1}{\tilde{f}_p^2}
\sum_q \left(1-\frac{m_q^2}{m_{\phi}^2}\right)^{3/2} \quad \quad \tilde{f}_p \equiv  \sum_q \frac{m_p}{m_q} f^{(p)}_{Tq}\,. 
\end{equation}
The second possibility is that they are proportional to the Yukawa couplings, $c_q = \tilde{c} \, \sqrt{2} \, m_q/v$, with the correlation becoming:
\begin{equation}
\sv_{v \rightarrow 0} = \sigma_p \cdot 3 \frac{(m_\phi+m_p)^2}{m_p^2} \frac{1}{\hat{f}_p^2}
\sum_q \frac{m_q^2}{m_p^2} \left(1-\frac{m_q^2}{m_{\phi}^2}\right)^{3/2} \quad \quad \hat{f}_p \equiv  \sum_q f^{(p)}_{Tq}\,. 
\label{eq:Yukcoupl}
\end{equation}
Within our standard choice of parameters $\tilde{f}_p = 18.5$ and $\hat{f}_p=0.375$; $m_\phi =10$~GeV and $\sigma_p = 2\cdot  10^{-41}$~cm$^2$, values compatible with CoGeNT data according to Ref.~\cite{Aalseth:2010vx}, gives in the first case $\sv \simeq 3 \cdot 10^{-30}$~cm$^3$~s$^{-1}$, i.e., a very small value most likely not  testable with indirect detection techniques, while in the second case $\sv \simeq 3 \cdot 10^{-26}$~cm$^3$~s$^{-1}$, the nominal value required for the thermal relic density of a WIMP to account for the dark matter in the Universe. 

\subsection{The antiproton source function for the DM component}\label{sec:DMprofile}

\begin{figure}[tbp]
\centering
\includegraphics[width=0.9\textwidth]{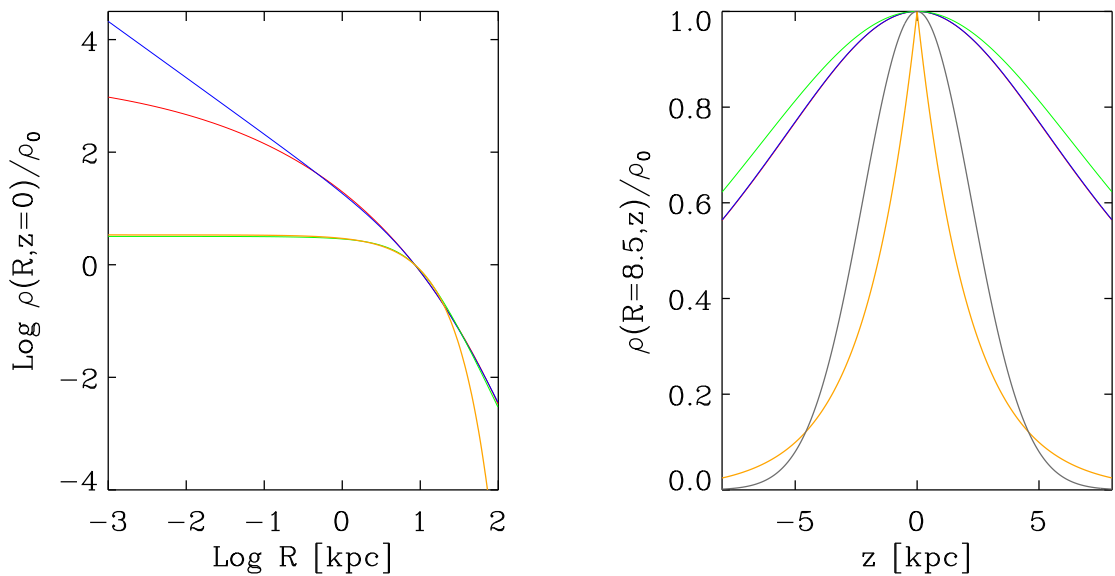}
\caption{DM density profiles versus the radial coordinate ({\it Left panel}: $z=0$) and vertical coordinate ({\it Right panel}: $R$ equal to the local galactocentric distance $R_\odot$) in a cylindrical frame. Both the spherical profiles (Einasto - red; NFW - blue; Burkert - green) and the two
dark disc profiles (Eq.~\ref{eq:DD1} - orange; Eq.~\ref{eq:DD2} - gray) are normalized to 1 at our position in the Galaxy.}
\label{fig:DMprofiles}
\end{figure}

The source function for the WIMP DM component scales with the number density of WIMP pairs in the Galaxy times the probability of annihilation and the antiproton yield per annihilation, namely it takes the form: 
\begin{equation} \label{eq:dmsource}
  Q_{\bar{p}}(\vec{r}, t,p) =  \frac{1}{2} \left(\frac{\rho_{\chi}(\vec{r})}{m_{\chi}}\right)^{2} \frac{dN_{\bar{p}}}{dE} \, \sv \,,
\end{equation}
where $m_\chi$ is the DM particle mass, $\sv$ the pair annihilation cross section in the limit of small relative velocity $v$ for the incoming particles, and ${dN_{\bar{p}}}/{dE}$ the antiproton emission spectrum. All these quantities are fixed once a specific WIMP DM candidate is selected, such as, e.g., within the three sample frameworks introduced above. The further ingredient one has to provide, independent of the specific WIMP model, is the spatial distribution of WIMPs in the Milky Way. In most of our analysis, we will assume that the DM density profile is spherically symmetric and takes the form:
\begin{equation}
  \rho_\chi(r)=\rho^{\prime} f\left(r/a_h\right)\,,
\label{nbody}
\end{equation}
where, as suggested from results of N-body simulations of hierarchical clustering, $f(x)$ is the function which sets the universal (or nearly universal) shape of dark matter halos, while $\rho^{\prime}$ and $a_h$ are a mass normalization and a length scale, usually given in terms of the virial mass $M_{vir}$ and a concentration parameter $c_{vir}$.
The dynamical constraints available for the Milky Way provide only weak discriminations among viable dark matter density profiles. We will consider three sample cases:  
the latest simulations favor the Einasto profile~\cite{Navarro:2004,Graham:2006}:
\begin{equation}
f_{E}(x) = \exp\left[-\frac{2}{\alpha_E} \left(x^{\alpha_E}-1\right)\right]\,, 
\label{eq:einasto}
\end{equation}
with the Einasto index $\alpha_E$ ranging about 0.1 to 0.25  (we take $\alpha_E=0.17$ as a reference value); we will also derive results for the profile originally proposed by Navarro, Frenk and White (NFW)~\cite{NFW}, i.e.,
\begin{equation}
f_{NFW}(x)=\frac{1}{x(1+x)^2} \,, 
\end{equation}
which is most often used in the dark matter studies. Finally we will consider the Burkert profile~\cite{Burkert:1995yz}: 
\begin{equation}
f_{B}(x)=\frac{1}{(1+x)(1+x^2)} \,, 
\end{equation}
in which the central enhancement in the dark matter profile predicted in the numerical simulations is totally erased, possibly as a backreaction of a baryon infall scenario with large exchange of angular momentum between the gas and dark matter particles, see, e.g.~\cite{ElZant:2003rp}.  A density profile with a constant core is also phenomenologically motivated since it reproduces better the gentle rise in the rotation curve at small radii which seems to be observed for many external galaxies, especially in the case of low-mass dark-matter-dominated low surface brightness  and dwarf galaxies \cite{Gentile:2004}. The free parameters in the three models are chosen following the analysis in Ref.~\cite{Catena:2009mf}, where a new study  on the problem of constructing mass models for the Milky Way was performed, comparing with a vast sample of dynamical observables for the Galaxy, including several recent results, and implementing a Bayesian approach to the parameter estimation based on a Markov chain Monte Carlo method. We adopt here profiles corresponding to the mean values in the resulting distributions, as specified in Table~\ref{halopar2}, having selected the local halo density $\rho_\chi(R_\odot)=0.4$~GeV~cm$^{-3}$ for all three models (it is convenient to compare different profiles using the same local normalization, and, in any case, such value of the local halo density is close to the best fit value for each of the three profiles). The conventions we used to define virial mass and concentration parameters are: $M_{vir}\equiv 4\pi/3 \Delta_{vir} \bar{\rho}_0\, R_{vir}^3$, with $\Delta_{vir}$ the virial overdensity as computed in Ref.~\cite{Bryan:1998}, $\bar{\rho}_0$ the mean background density and $R_{vir}$ the virial radius; and $c_{vir} \equiv R_{vir}/r_{-2}$, with $r_{-2}$ the radius at which the effective logarithmic slope of the profile is $-2$.
Note finally that the value of concentration parameters in the Table refer to a fit  of the profile to the Galaxy and not to the dark matter density before the baryon infall; hence a direct comparison with values found with numerical simulations for the dark matter component only (which, in general, are lower for Milky Way size halos) is not straightforward. The shape of the three spherical halo models is shown in Fig.~\ref{fig:DMprofiles}; with our choice of parameters, the Einasto and the NFW profile trace each other down to fairly small radii, while the cored Burkert profiles shows a more evident departure from the others.
\begin{table}[tbp]
\centering
\begin{tabular}{|c|c|c|c|}
\hline
Parameter & Einasto & NFW & Burkert  \\
\hline \hline
$M_{vir}$ [10$^{12}$~M$_{\odot}$] & 1.3 & 1.5 & 1.3 \\
$c_{vir}$ & 18.0 & 20.0 & 18.5 \\
$\alpha_E$ & 0.22 & - & - \\
$\rho_\chi(R_\odot)$ [ GeV~cm$^{-3}$] & 0.4 & 0.4 & 0.4 \\
$a_h$ [ kpc ] & 15.7 & 14.8 & 10.0 \\
\hline 
\end{tabular}
\caption{Parameters defining the dark matter halo profiles implemented for this analysis. The value of the local halo density $\rho_\chi(R_\odot)$ and the halo scale factor $a_h$ are given here for 
reference, being a derived quantity if we adopt as mass scale parameter the virial mass $M_{vir}$ and length scale the concentration parameter $c_{vir}$. \label{halopar2}}
\end{table}
Recently, cosmological simulations including baryons \cite{Read08,Read09,Pato10} have suggested the existence of a dark disk substructure within CDM halos, with a characteristic scale height of the order of 1~kpc. Would a dark disc be present in the Milky Way, it would have an impact on the WIMP antiproton source function; we will discuss this effect using two alternative parameterizations for the dark disk (DD) profile, differing only in vertical shape, namely~\cite{Read09}:
\begin{equation}\label{eq:DD1}
  \rho_{\chi, {\rm DD}}(R,z)=\rho_{0,\rm DD} \, {\rm e}^{\left(1.68(R_\odot-R)/R_{H}\right)} {\rm e}^{(-0.693 z/z_{H})}
\end{equation}
and~\cite{Read09}:
\begin{equation}\label{eq:DD2}
  \rho_{\chi, {\rm DD}}(R,z)=\rho_{0,\rm DD} \,{\rm e}^{\left(1.68 (R_\odot-R)/R_{H}\right)} {\rm e}^{\left(-(0.477 z/z_{H})^{2}\right)}\,,
\end{equation}
with $z_{H} = 1.5$~kpc and $R_{H} =11.7$~kpc. An alternative parameterization of the vertical profile in terms of the inverse of the square the hyperbolic cosine, also given in Ref.~\cite{Read09}, would give essentially the same results as Eq.~\ref{eq:DD2}. Thicker disks ($z_{H} =2.8$~kpc and $R_{H} =12.6$~kpc) have also been suggested using HI data~\cite{Kalberla07}; however, as explained below, since the effect of a thick disk can be mimicked by changing the height of the CR diffusion zone to about $z_{H}$, we will not consider this model.

When including a dark disk, there are two source functions contributing to the DM induced antiproton flux and we need to fix the relative normalization. Reference~\cite{Read09}, has suggested as an upper estimate on the local DM density in the dark disc compared to the local DM density from the spherical halo component to be $\rho_{0,\rm DD}/\rho_\chi(R_\odot) \approx 1.5$. For simplicity, in our simulations with a dark disk component, we will assume $\rho_{0,\rm DD}/\rho_\chi(R_\odot) = 1$, which is close to the maximal dark disk contribution we could have. 
We also still keep fixed the total local DM density (namely $\rho_{0,\rm DD}+\rho_\chi(R_\odot)$) to be 0.4~GeV~cm$^{-3}$, since, for comparison, it is still convenient to have same local normalization of the DM source function.
Such a choice decreases the total dark matter mass included within $R_\odot$ by $\approx 1/3$, see the plot of radial and vertical profiles in Fig.~\ref{fig:DMprofiles}, and thus would have, e.g., an effect on the star circular velocity at $R=R_\odot$, one of the pieces of information which has also been used in getting the value of 0.4~GeV~cm$^{-3}$ in \cite{Catena:2009mf} (see also \cite{Salucci:2010qr}).
Thus in estimating the local value of the DM density an analysis including the possible presence of a dark disk would be necessary, which is however beyond the scope of this paper.

\section[]{Selection of CR propagation models: signal versus background}\label{sec:cr_models}

The propagation of CRs in the Galaxy is governed by the following transport equation \cite{Berezinskii90}:
\begin{eqnarray}
\frac{\partial N_i}{\partial t} &-& {\bm \nabla}\cdot \left( D\,{\bm \nabla}
-\bm{v}_{c}\right)N_{i} + \frac{\partial}{\partial p} \left(\dot{p}-\frac{p}{3}\bm{\nabla}\cdot\bm{v}_{c}\right) N_i -\frac{\partial}{\partial p} p^2 D_{pp}
\frac{\partial}{\partial p} \frac{N_i}{p^2} =  \nonumber \\
&=&  Q_{i}(p,r,z) + \sum_{j>i}c\beta n_{\rm gas}(r,z)
\sigma_{ji}N_{j} -  c\beta n_{\rm gas}\sigma^{\rm in}_{i}(E_{k})N_{i}\;,
\label{eq:diffusion_equation}
 \end{eqnarray}
in which $N_i(p,r,z)$ is the number density of the $i$-th atomic species; $p$ is its momentum; $\beta$ the particle velocity in units of the speed of light $c$; $\sigma_i^{\rm in}$ is the total inelastic cross section onto the interstellar medium (ISM) gas, whose density is $n_{\rm gas}$; $\sigma_{ij}$ is the production cross section of a nuclear species $j$ by the fragmentation of the $i$-th one; $D$ is the spatial diffusion coefficient; $\bm{v}_{c}$ is the convection velocity. 

The diffusion coefficient $D$ is assumed to be of the form:
\begin{equation}
\label{eq:diff_coeff}
 D(\rho,R,z) = D_0 ~\beta^\eta g(R,z) \left(\frac{\rho}{\rho_0}\right)^\delta \;,
 \end{equation}
with $\rho \equiv p\beta c/(Ze)$ being the rigidity of the nucleus of charge $Z$ and momentum $p$, $g(R,z)$ describing the spatial dependence (in cylindrical coordinates) of $D$, and $\eta$ controlling essentially the low energy behavior of $D$. 
While one would expect  $\eta=1$ as the most natural dependence of diffusion on the particle speed, several effects may give rise to a different effective behavior. 
For example, it should be taken into account that diffusion may actually be enhanced at low energies due to the back reaction of CRs on the magneto-hydrodynamic waves.
In a dedicated analysis of that effect, a low-energy increase of $D$  was found \cite{Ptuskin:2005ax}. While such a behavior cannot be represented as a simple function of $\beta$ and $\rho$,  an effective value of $\eta$ may, nevertheless, be found which allows to fit low energy data. 
Clearly, the required value of $\eta$ depends on the details of the model. For example in \cite{Ptuskin:2005ax} $\eta = - 3$ was found, while the authors of \cite{Maurin10} found $\eta = -1.3$, in both cases for models with $\delta = 0.5$ (but rather different values of other parameters). In \cite{DiBernardo10}  $\eta = -0.4$ was found to allow a rather good fit of low energy nuclear data for models with low reacceleration and $\delta \simeq 0.5$. Here, where not differently stated, we tune $\eta$, as other parameters, by minimizing the $\chi^2$ of the model against B/C and  proton data (see below).

The spatial behavior of $D(\rho,R,z)$ is largely unknown. In the following, we will assume that the function $g(R,z)$ can be factorized as
\begin{equation}
g(R,z) = G(R) e^{|z|/z_{t}}\;, \label{D_vert}
\end{equation}
and we will set $G(R) = 1$ whenever we do not explicitly mention a different radial dependence. The vertical dependence of the diffusion coefficient
is assumed to be exponential with scale height $z_{t}$, in correlation with the scaling of magnetic fields in the Galaxy. Notice that this assumption is opposed to most analyses in the literature which assume instead that $D$ does not depend on $z$, however our results either for standard and exotic components depend mildly on this choice and we do not discuss further. We set the vertical size of the propagation box as: $z_{\rm max} = 2 \times z_{t}$.

The last term on the left-hand side~of Eq.~(\ref{eq:diffusion_equation}) describes diffusive reacceleration of CRs in the turbulent galactic magnetic field.  In agreement with the quasilinear theory we assume the diffusion coefficient in momentum space $D_{pp}$ to be related to the spatial diffusion coefficient by the relationship (see e.g.,~\cite{Berezinskii90}) 
\begin{equation}
D_{pp} = \frac{4}{3 \delta (4 - \delta^2)(4 - \delta)} \frac{v_A^2~p^2}{D}\;,
\label{eq:dpp}
\end{equation}
where  $v_A$ is the Alfv\'en velocity. Here we assume that diffusive reacceleration takes place in the entire diffusive halo. 

For the CRs generated by standard astrophysical sources, $Q^{i}(p,r,z)$ will describe the distribution and injection spectrum of SNRs, which we parametrize as
\begin{equation}
Q_{i}(E_{k},r,z) =  f_S(r,z)\  q_{0,i}\ \left(\frac{\rho(E_{k})}{\rho_0}\right)^{- \gamma_{i}} \;,
\label{eq:source}
\end{equation}
In this paper we assume the same source spectral index $\gamma_{i} = \gamma$ for all nuclear species unless differently stated. 
We require the source spatial distribution $f_S(r,z)$ to trace that of Galactic supernova remnants inferred from pulsars and stellar catalogues as given in \cite{Ferriere:2001rg}.  We checked that other distributions, among those usually adopted in the literature, do not affect significantly our results. 
For the case of DM annihilations, the source is given above in Eq.~(\ref{eq:dmsource}) where the antiproton yield per annihilation ${dN_{\bar{p}}}/{dE}$
is obtained interfacing the numerical code with the \ds\ package~\cite{Gondolo:2004sc}, 
in turn linking to simulations with the \pythia\ Monte Carlo, except for the heavy WIMPs
models for which tables provided by~\cite{Ciafaloni:2010ti} are used instead.  

Secondary antiprotons are generated in the interaction of primary CRs with the interstellar gas.
The ISM gas is composed mainly by molecular, atomic and ionized hydrogen (respectively, H2, HI and HII). Here we adopt the same distributions as in \cite{Strong98,Evoli:2008dv}. Following \cite{Asplund:2006} we take the He/H numerical fraction in the ISM to be 0.11. We have tested that different models for the gas distribution (i.e.,~\cite{Nakanishi:2003,Nakanishi:2006}) affects marginally the fitted model parameters and hence the predicted antiproton spectra. 

The diffusion equation offers just an effective description of the CR transport in the Galaxy. The main parameters determining the propagated distribution and spectrum of CR nuclei are the normalization of the diffusion coefficient $D_{0}$, its vertical scale $z_{t}$ and its rigidity slope $\delta$, the Alfv\'en velocity $v_{A}$ and the convection velocity $\bm{v}_{c}(R,z)$. Presently available observations of secondary/primary ratios, like the B/C, or unstable/stable ratios, like $^{10}$Be/$^{9}$Be allow to determine such parameters only up to large uncertainties (see \cite{DiBernardo10} for a reference list of the experimental data). Moreover, secondary-to-primary ratios are sensitive only to the ratio $D_{0}/z_{t}$, while unstable-to-stable ratios, that are somewhat more sensitive to $D_{0}$ and $z_{t}$ separately and can therefore break the degeneracy, suffer from large experimental uncertainties. Therefore, the half-height of the diffusion region $z_{t}$ is poorly constrained by CR nuclei observations. Radio and $\gamma$-ray observations are more sensitive to $z_t$ and seem to disfavor small values $z_t \lesssim 1~\kpc$ (see e.g.,~the recent works~\cite{Bringmann:2011py,Cholis:2011un}). To place an upper bound on $z_t$ requires instead more careful analyses. However, the parameter $z_{t}$ might affect significantly the flux expected from DM sources, as they are also distributed in the galactic halo. Also the antiproton fraction reaching the Earth from the galactic center region depends strongly on $z_t$. For this reasons, we consider 5 different reference models, encompassing a range of possible propagation regimes, which we summarize in Table~\ref{tab:models}: Models KRA, THN and THK assume Kraichnan type turbulence ($\delta = 0.5$) but differ in the adopted height of the diffusion zone in order to probe the effect of varying this parameter on the ${\bar p}$ flux; the KOL model assumes instead Kolmogorov turbulence ($\delta = 0.33$); the CON model considers convective effects.
\begin{table}[tbp]
\caption{We report here the main parameters of the reference CR propagation models used in this work. The KOL and CON models have a break in rigidity the nuclei source spectra $\gamma$ at respectively, 11 GV and 9 GV. The modulation potential $\Phi$ refers to the fit of proton PAMELA data only.}
\begin{tabular}{|c|c|c|c|c|c|c|c|c|c|c|c|c|}
\hline
{\bf Model} & $z_t (\kpc)$ & $\delta$ & $D_0 (10^{28}\cm^{2}/\s)$ & $\eta$ & $v_A (\km/\s)$ & $\gamma$ & $dv_c/dz (\km/\s/\kpc)$ & $\chi^{2}_{B/C}$ & $\chi^{2}_{p}$ & $\Phi~(\rm GV)$ & $\chi^{2}_{\bar{p}}$ & Color in Fig.s \\
\hline
{\bf $KRA$} & $4$ & $0.50$ & $2.64$ & $-0.39$ & $14.2$ & $2.35$ & $0$ & $0.6$ & $0.47$ & $0.67$ & 0.59 & \textcolor{red}{Red} \\
\hline
{\bf $KOL$} & $4$ & $0.33$ & $4.46$ & $1.$ & $36.$ & $1.78/2.45$  & $0$ & $0.4$ & $0.3$ & $0.36$ & 1.84 & \textcolor{blue}{Blue} \\
\hline
{\bf $THN$} & $0.5$ & $0.50$ & $0.31$ & $-0.27$ & $11.6$ & $2.35$  & $0$ & $0.7$ & $0.46$ & $0.70$ & 0.73 & \textcolor{green}{Green} \\
\hline
{\bf $THK$} & $10$ & $0.50$ & $4.75$ & $-0.15$ & $14.1$ & $2.35$  & $0$ & $0.7$ & $0.55$ & $0.69$ & 0.62 & \textcolor{orange}{Orange} \\
\hline
{\bf $CON$} & $4$ & $0.6$ & $0.97$ & $1.$ & $38.1$ & $1.62/2.35$  & $50$ & $0.4$ & $0.53$ & $0.21$  & 1.32 &\textcolor{black}{Gray} \\
\hline
\end{tabular}
\label{tab:models}
\end{table}
All these models are chosen in such a way as to minimize the combined $\chi^2$ against B/C and the proton spectrum data under the requirement to get  $\chi^2 < 1$ for each of those channels. An accurate modeling of proton data is crucial since protons are the main primaries of secondary antiprotons.
For the first time in the context of secondary antiproton computations, the proton spectrum is fitted against the high precision data recently released by the PAMELA collaboration \cite{Adriani:2011cu}. We also checked that the $^4$He spectrum measured by the same experiment is reproduced by each of those models.
The fits are performed minimizing the $\chi^2$ in the multidimensional parameter space defined by $D_0$, $\eta$, the Alfv\'en velocity $v_A$,  the proton and nuclei spectral indices $\gamma_{i}$, the solar modulation potentials $\Phi$.
For some models a spectral break has to be introduced in the source proton spectrum in order to achieve an acceptable fit ($\chi^2_p < 1$) of proton data (see below). For those models the spectral indexes below/above the break and the break rigidity are also fitted.    

\begin{figure}[tbp]
\centering
\includegraphics[width=0.45\textwidth]{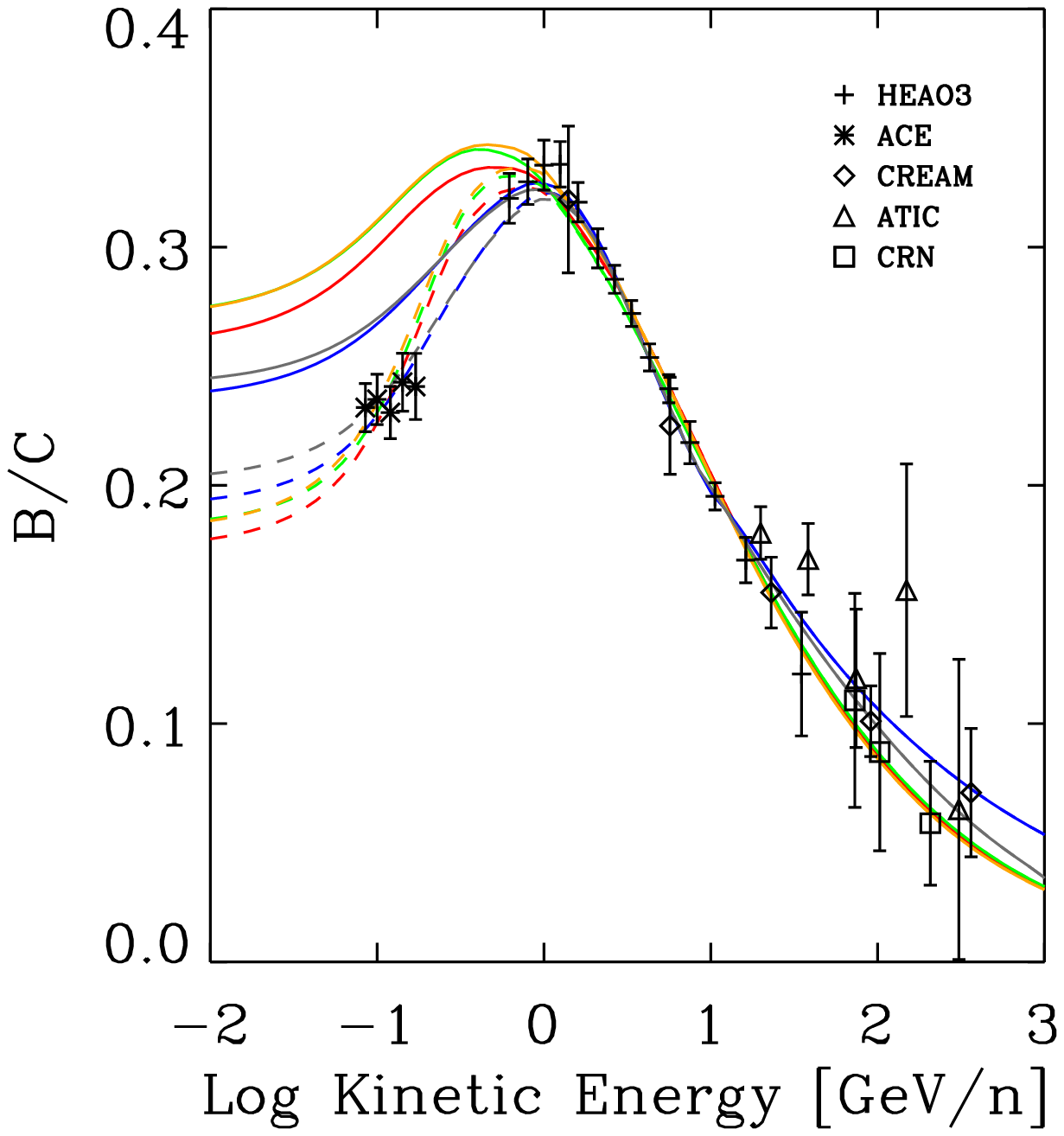}
\includegraphics[width=0.45\textwidth]{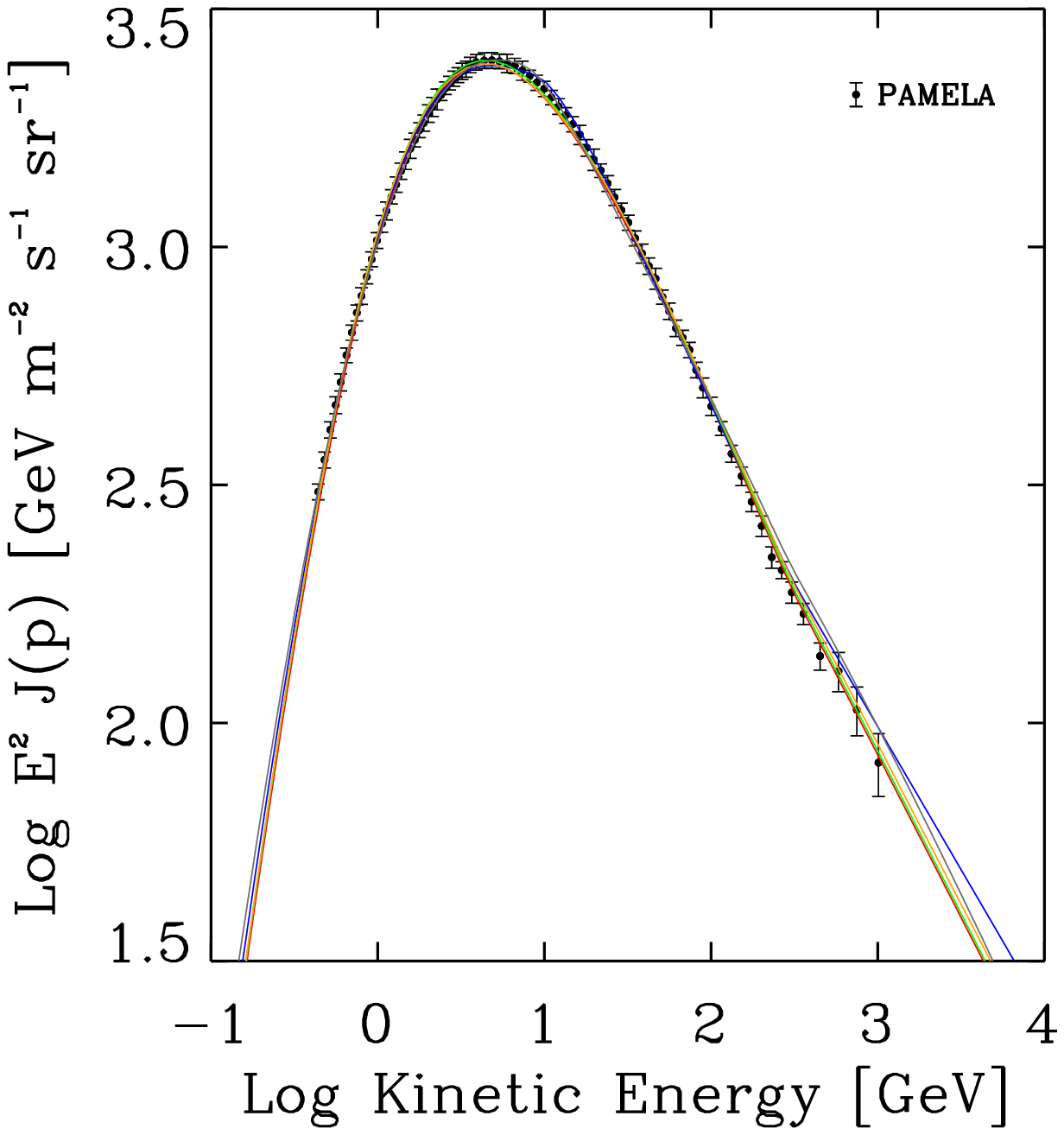}
\caption{{\it Left panel}: Comparison of reference models with B/C data (solid: modulated with a potential of 550 MV, dashed: with a potential of 300 MV or 220 MV, see Sec.~\ref{sec:cr_models}). KRA (red), KOL (blue), THN (green), THK (orange), CON (gray), see Table~\ref{tab:models}. {\it Right panel}: The proton spectrum computed for the same models modulated with a potential given in Table~\ref{tab:models} are compared with PAMELA data \cite{Adriani:2011cu}.}
\label{fig:bcmodels}
\end{figure}

\begin{figure}[tbp]
\centering
\includegraphics[width=0.45\textwidth]{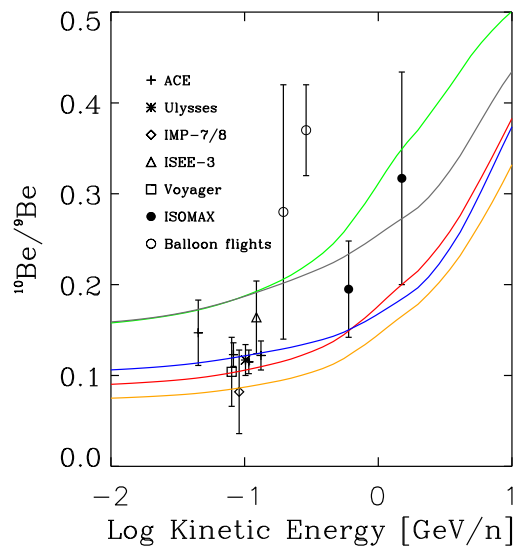}
\caption{The $^{10}$Be/$^9$Be ratio computed for the reference models in Table~\ref{tab:models}, modulated with a potential $\Phi = 400$ MV. The color coding is the same as in Fig.~\ref{fig:bcmodels}. }
\label{fig:be_models} 
\end{figure}

The propagation equation is solved with the public available \dragon\ code~\cite{Evoli:2008dv}, implementing a numerical solution which assumes cylindrical symmetry and a  stationary state.
In Fig.~\ref{fig:bcmodels} spectra for our selected sample of models, as obtained after the fitting procedure, are plotted against the B/C ratio, and the most relevant proton data. Because of their strong scatter, presently available $^{10}$Be/$^{9}$Be data cannot reliably be used in a statistical fit. We checked, however, that all models in Table~\ref{tab:models} are roughly compatible with those data (see Fig.~\ref{fig:be_models}). 

For the KRA, THN and THK models (same Kraichnan type turbulence, different values of $z_t$) relatively low values of the Alfv\'en velocity ( $10 \div 15~ \km/\s$) and negative values of $\eta$ provide the best fit of the data. No spectral break is required to reproduce proton data. The KRA model is actually very similar to the best fit model found in~\cite{DiBernardo10}. The KOL model, in agreement with previous findings \cite{Strong98}, requires a larger $v_{A}$. 
While that model allows to reproduce the data with the conventional value $\eta = 1$, it requires to introduce an {\it ad hoc} break in the primary proton source spectrum in order to reproduce the observed proton spectrum. This is a well-known prescription which has to be imposed to propagation models with strong reacceleration (see {\it e.g.} \cite{Strong04}). One should also notice that strong reacceleration models might be in contrast with synchrotron observations \cite{Jaffe:2011qw}.

The models mentioned above do not account for the presence of convection.  However, stellar winds can be very effective in removing CRs from the galactic plane, hence their presence may affect significantly CR propagation and DM originated fluxes.  The Galactic diffuse soft X-ray emission observed by ROSAT has been interpreted with the presence of a strong Galactic wind \cite{Everett07,Breitschwerdt08}.  
Therefore, we build the CON model in order to probe the effects of convection. We assumed $z_t = 4~\kpc$, fixed the convective speed to zero on the Galactic plane and assumed a uniform gradient in the $z$ direction, directed outwards the galactic plane, $dv_c/dz = 50$~km/s/kpc, to have a velocity of several hundreds km/s at the halo hedge.  
For this model, we fix $\eta=1$ and all other parameters, including $\delta$, are then fitted against the data.  
The best fit value of $\delta= 0.6$ is larger than the other models as expected in order to compensate the low energy CR depletion produced by convection. 

Solar modulation has to be taken into account for a correct modeling of the CR spectra below few $~\GeV/{\rm n}$. Similarly to what done in most related papers, we treat solar modulation in the "force field" charge independent scenario \cite{Gleeson68}. As it is well known, in this scenario modulation can be parametrized in terms of an {\it effective potential} $\Phi$. Since $\Phi$ is a model dependent parameter, for each model we also fit it against the data. For the B/C we fix the modulation potential to the value of $550~$MV for data above $\sim$~GeV, ACE data need instead a rescaling of the modulation potential as done in previous works. We find that a modulation potential of $300$~MV for the KOL and CON models and $220$~MV for the KRA, THN e THK models are in good agreement with the data.

\begin{figure}[tbp]
\centering
\includegraphics[width=0.9\textwidth]{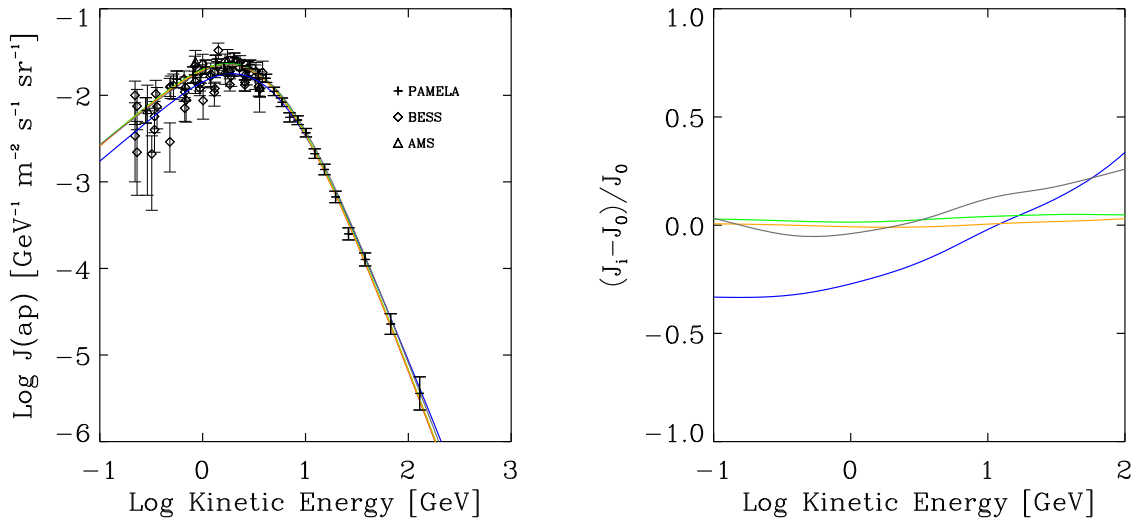}
\caption{{\it Left panel}: Comparison of the local spectrum of secondary antiprotons for different propagation models (modulated with a potential as given in Table~\ref{tab:models}). {\it Right panel}: Fractional ratio between the different local spectrum and the KRA model.}
\label{fig:apsec}
\end{figure}

Low energy proton data are the most sensitive to $\Phi$. When comparing against PAMELA ${\bar p}$ data, which provide the strongest constrains on DM models,
we use the PAMELA proton data, over their entire energy range, to fit the parameter $\Phi$ (see Table~\ref{tab:models}). In some cases we will use other antiproton datasets (see Fig.~\ref{fig:apsec}) and to properly take into account the effect of modulation we refit the modulation potential against the proton flux as measured from the same experiment in the same solar cycle period. The antiproton and proton data are taken from \cite{Asaoka:2001fv} for BESS and from \cite{Aguilar:2002ad} for the AMS-01 experiment.
\subsection{Secondary antiprotons}

As we discussed in the introduction, secondary antiprotons are an unavoidable byproduct of CR propagation and are the major background for indirect DM searches.
We use \dragon\ to determine the secondary antiproton spectrum for each model in Table~\ref{tab:models}. Our approach is the same followed in \cite{DiBernardo10} (to which we address the reader for details) and it is similar to that discussed in several previous papers \cite{Donato01,Moskalenko02}.
Our analysis accounts for the scattering $p-p_{\rm ISM}$,  $p-^4{\rm He}_{\rm ISM}$, $^4{\rm He}-p_{\rm ISM}$ and $^4{\rm He}-^4{\rm He}_{\rm ISM}$ and for annihilation and inelastic, nonannihilating, scattering of $\bar{p}$ onto the ISM gas. The contribution of heavier CR and ISM nuclei is negligible.  
Based on the data from ISR STAR and ALICE experiments \cite{Alper1975237, :2008ez, Aamodt:2010dx} there is an energy dependent uncertainty up to $\pm 9$\% on the multiplicity ratio of produced antiprotons relative to the produced protons; propagating such uncertainty would have an impact on our final results within a few \%. Notice however that this is a minimum level of uncertainty one should include on the antiproton production cross section.
Reference~\cite{Donato01} has evaluated the nuclear physics uncertainties by computing all the relevant cross sections using the Monte Carlo program {\sffamily DTUNUC}. Their results suggest 25\% uncertainty in the propagated flux from the nuclear physics, which is below the 40\% uncertainty in the antiproton prediction that~\cite{1998ApJ...499..250S} has suggested by comparing the difference between the results for p-p collisions, of the {\sffamily DTUNUC} Monte Carlo simulation with those from the cross section parametrizations of~\cite{0305-4616-9-10-015} and of~\cite{1987SSRv...46...31S}.

We find that all models, which are built to reproduce the B/C data, provide a good fit also of the antiproton measured spectrum above a few GeV.
At lower energies the KOL model underproduces ${\bar p}$ (see  Fig.~\ref{fig:apdm}). This is a well known feature of models with strong reacceleration (see e.g.,~\cite{DiBernardo10}).  
From the right panel of Fig.~\ref{fig:apdm} we see that the maximal scatter on the secondary proton spectrum amounts to $\pm 30~\%$ in the $0.1 \div 10^2~\GeV$ energy range which turns into significant uncertainties on the room possibly left for a DM ${\bar p}$ component.  
\subsection{Antiprotons from WIMP annihilations}
\begin{figure}[tbp]
\centering
\includegraphics[width=0.9\textwidth]{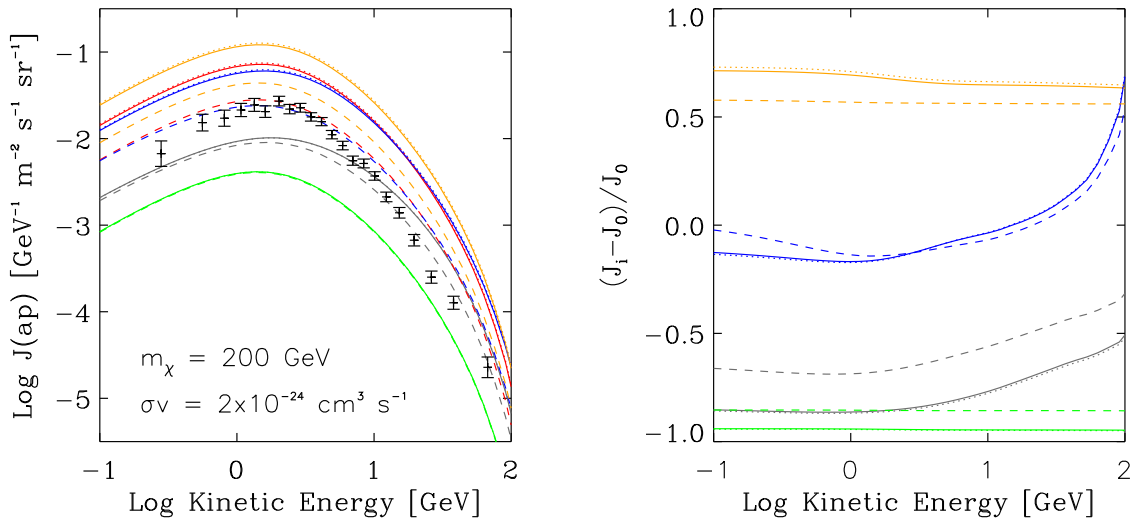}
\caption{{\it Left panel:} Comparison of the local spectrum of antiprotons from 200 GeV Wino DM ($\sv$ = $2\times 10^{-24} {\rm cm}^{3} {\rm s}^{-1}$) for different propagation models (the color coding is the same as in Fig.~\ref{fig:bcmodels}), assuming a modulation potential as given in Table~\ref{tab:models} and the three spherical halo model profiles introduced in Table~\ref{halopar2} (solid: Einasto profile, dotted: NFW, dashed: Burkert). {\it Right panel}: Fraction ratio between the different local spectrum and the KRA model. In some cases solid and dotted curves coincide.}
\label{fig:apdm}
\end{figure}

For the same set of diffusion models we have just introduced, in Fig.~\ref{fig:apdm} we show the predictions obtained with \dragon\ for a first sample WIMP model, a pure Wino with mass equal to 200 GeV, annihilating in pairs into W-bosons with a cross section of $\sv = 2\times 10^{-24}\, {\rm cm}^3 {\rm s}^{-1}$. For each propagation model results are shown for the three spherical DM distributions introduced in Table~\ref{halopar2}.
As evident from the plot,  the antiproton flux from WIMP DM annihilations is much more dependent upon the propagation model than the secondary component. Predictions are also clearly sensitive to how the source function changes away from the local neighborhood (the three halo profiles are normalized in the same way at the local galactocentric distance), with the local antiproton flux being in some of the models significantly larger for DM density profiles which are enhanced in the galactic center region. Summing the two effects, the spread in the predictions for this single DM candidate is larger than a factor of 40, to be compared to the 30\% spread at low energy in the secondary component (also compare the left hand side of Fig.~\ref{fig:apsec} and Fig.~\ref{fig:apdm}). The range of uncertainty found here is comparable to what has been found in previous studies in the literature \cite{Donato01,Donato:2008jk} and brings in a number of questions that we are going to address in detail in the next section discussing locality or nonlocality issues. 

\section{Locality tests} \label{sec:local}

To discuss the origin of the discrepancies in the ratio between the signal from DM annihilations and the background from secondary production within the set of propagation models and dark matter distributions we are considering, it is important to study the dependence of the antiproton flux at our location in the Galaxy as a function of the position where the antiprotons are generated in the two cases. 

We start by testing a close analogue in our numerical solution of what would be the local response in the ${\bar p}$ flux to a point DM source of ${\bar p}$ if we would implement a solution of the propagation equation with the Green function method. Since we are working with a numerical code which assumes cylindrical symmetry and finite step size in radial ($\Delta R$) and vertical ($\Delta z$) directions, we define a ``ringlike" source function on our grid:
\begin{eqnarray}
Q_{\bar{p}}(R,z; \bar{R}, \bar{z}) &\propto& \frac{1}{R \Delta R \Delta z}\;, \qquad \bar{R}-\Delta R/2 < R < \bar{R} + \Delta R/2 \quad \bar{z}-\Delta z/2 < z < \bar{z} + \Delta z/2 \\
\nonumber & & 0 \qquad \qquad \qquad{\rm otherwise}
\end{eqnarray}
i.e.,~a source with ring shape and parallel to the Galactic plane, which we will normalize setting to 1 the flux for a ``ringlike" source of $R=R_\odot$. All results for DM components shown in this section are obtained assuming the 200 GeV Wino model introduced above. However, since the effect of energy redistributions are marginal for antiprotons along propagation, the results we present in this section are independent of this choice. 

\begin{figure}[tbp]
\centering
\includegraphics[width=\textwidth]{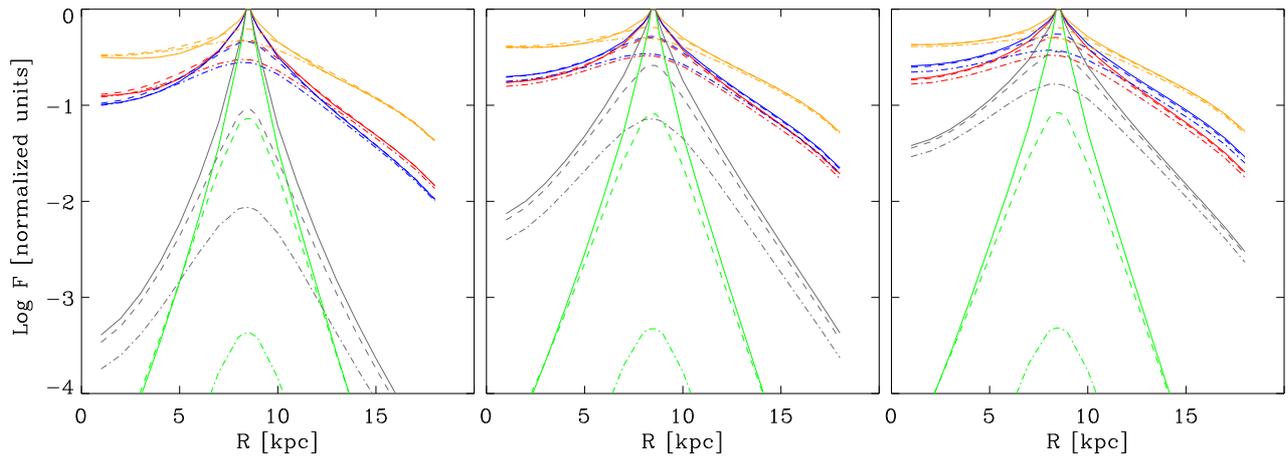}
\caption{Flux observed at Earth for a ``ringlike" source located at distance $R$ from the GC and $z=0$ (solid line), $z=1$ (dashed), $z=2$ (dashed-dotted),
normalized to the flux for $R=R_\odot$ and $z=0$. From left to right the plots are for propagated $E_{k} = 1$, $10$, $100$ GeV. Color of lines represent different propagation models as in Fig.~\ref{fig:bcmodels}.} 
\label{dmpsource}
\end{figure}

In Fig.~\ref{dmpsource} we plot the response on the local antiproton flux to a DM source located at the galactocentric distance $R$ and vertical height $z$, for three different values of the kinetic energy of the locally observed (propagated) ${\bar p}$, $E_{k}= $ 1, 10, 100~GeV. Remarkably, the relevance of distant sources is very different for different propagation models which all reproduce the B/C and other CR nuclear data.  In particular, in the THN (green lines) and CON (grey lines) models, which are characterized by a small normalization of the diffusion coefficient, the relative ${\bar p}$ flux decreases rapidly 
with the source distance. 
For instance at $E_{k}= $ 10 GeV, the $\bar{p}$ flux arriving from $R=5$~kpc, is suppressed by a factor of 100 compared to the local flux in the THN model ($z_{t}=0.5$~kpc), a factor of 8 in the KRA case ($z_{t}=4$~kpc) and only a factor of 5 in the THK model ($z_{t}=10$~kpc).
This is expected since the THK model has the thickest diffusive halo size and the largest $D_0$, giving therefore the largest contribution from distant sources. 
In the convective model, instead, although we assumed the same halo thickness $z_{t}=4$~kpc as in the KRA and KOL models, the contribution of the ring source depends strongly on its position relative to ours. Again this is clear, as convection makes particles escape faster away from the disk, as does a smaller value of $z_{t}$.  Concerning the dependence of the ${\bar p}$ flux on the vertical position of the source, it is significant for small radial coordinates $R\lesssim 5~\kpc$, because the diffusion distance from there to the observation point at $R=8.5~\kpc$ and $z=0$ increases significantly with $z$.
We also notice that as we increase the distance $z$ of the source from the galactic plane, (see solid vs dashed vs dotted-dashed lines of Fig.~\ref{dmpsource}), the drop of the $\bar{p}$ flux relative to $R=8.5$~kpc is smoother.
Since we normalize to the flux at $R=8.5$~kpc and $z=0$~kpc from a source at the same position, and the diffusion coefficient increases exponentially with $z$ (as given in Eq.~\ref{D_vert}) a significant fraction of injected $\bar{p}$s at $z=1, \; 2$~kpc escapes before reaching  $z=0$; e.g., for injected $\bar{p}$ at $z=2$~kpc, $R=8$~kpc and $E_{k}=10$~GeV, $\simeq 50 \%$ of the $\bar{p}$s escape in the thick halo model THK, $\simeq 80 \%$  in the KRA (intermediate halo) model and $\simeq 95 \%$ in the THN (thin halo) model\footnote{In this case, for the THN model our simulation extended to a height of 3 kpc away from the disk.}. 
We also note that, differently from the case of $e^{\pm}$, in the antiproton (proton) case the diffusion timescales (escape times) are typically much smaller than the energy loss timescales ($\sim  E/(dE/dt)$). Within our models where the diffusion coefficient scales as $D \sim E^{\delta}$ with $\delta > 0$, higher energy CRs propagate via diffusion to greater distances, which explains why the 100~GeV $\bar{p}$ fluxes are less local compared to the 1~GeV $\bar{p}$ fluxes.
\begin{figure}[tbp]
\centering
\includegraphics[width=1\textwidth]{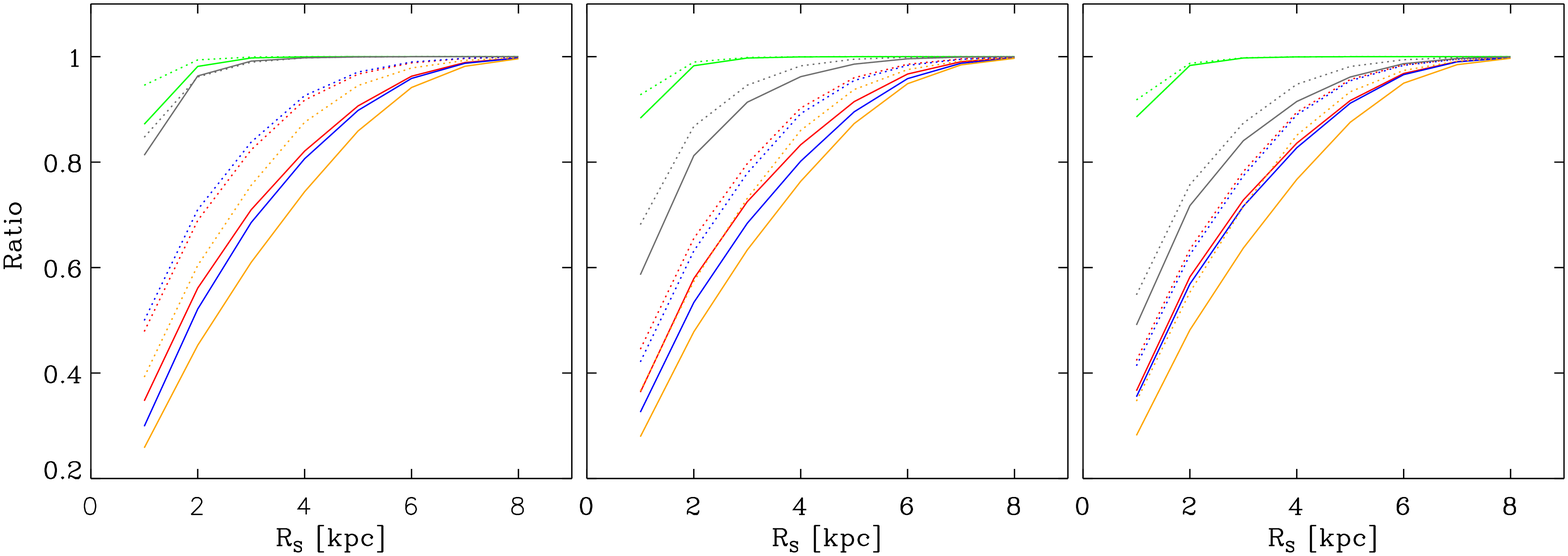} 
\includegraphics[width=1\textwidth]{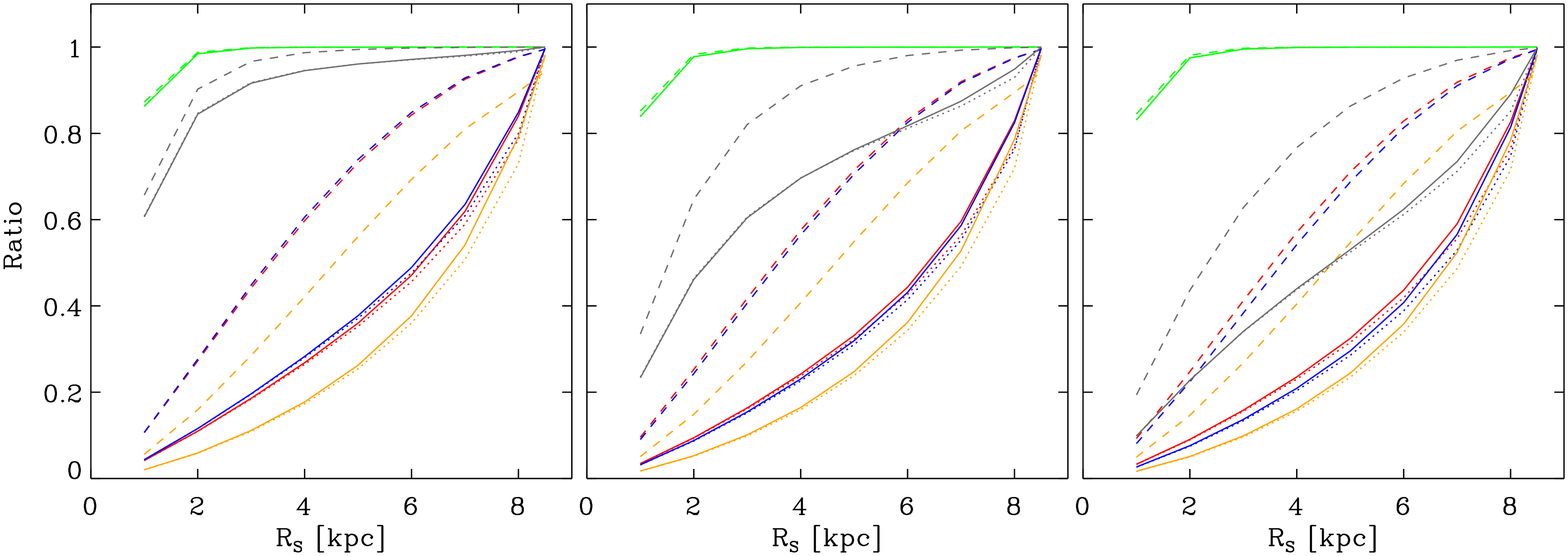}
\caption{Ratio of the local flux obtained considering sources with distance smaller than $R_{S}$ to that obtained with $R_S = \infty$: up) primary protons (solid line) and secondary antiprotons (dotted); down) antiprotons from DM (solid: Einasto, dotted: NFW, dashed: Burkert). From left to right the plots are for $E_{k} = 1$, $10$, $100$ GeV. Color code is the same as in Fig.~\ref{fig:bcmodels}.} 
\label{fig:test_source}
\end{figure}
In Fig.~\ref{fig:test_source} we introduce another more quantitative locality test by showing the contribution to the local fluxes given by sources located within a torus with axis at the galactic center and perpendicular to the Galactic plane, with major radius equal to our galactocentric distance $R_\odot$ and minor radius (radius of the tube) equal to the parameter $R_S$. For $R_S = R_\odot$ essentially the whole source function is included in the computation; we show results as normalized to this case. In the top panels we present our results for CR protons injected by SNRs and secondary antiprotons produced by CR spallation, while in lower panels those for DM antiprotons computed for the Wino model and the three spherically symmetric density profiles. As expected also from Fig.~\ref{dmpsource}, we see that for the THN and CON models the CR proton and secondary antiproton fluxes reaching the Earth are produced only within 3 kpc from the Earth;  for the other propagation models almost 90\% of the local flux is produced by SNRs or primary interactions within 6 kpc.  

It is again evident that the antiproton flux from DM annihilations is considerably more affected by the propagation parameters. For the very thin model THN it is very local irrespective of the DM halo profile and the energy of the detected antiproton. For the convective model CON, the relative contribution of local DM sources is still dominant, especially for the cored Burkert DM profile and at low energies. Remarkably, the DM distribution towards the Galactic Center has little effect in the CON model. For the other models the contribution of annihilations in regions close to the Galactic center can be very large and is indeed the dominant one for dark matter density profiles which are peaked towards the Galactic center (the annihilation rate is proportional to $\rho^2$). One can also see small differences between the Einasto and NFW profiles, which, as one can see in Fig.~\ref{fig:DMprofiles}, have sizable differences only for $r \lesssim 100$~pc.

A comparable analysis was already performed for semi analytic models in \cite{Taillet:2002ub,Maurin:2002uc}. Even if the models considered in this paper are not directly comparable with their setups, our results are compatible with their findings for ordinary sources and we confirm that halo height is the most important parameter in determining the locality of exotic contributions.

\section[]{Limits on DM models from antiproton data} \label{sec:constraints}

Since the $\bar{p}$ produced in $pp$ and $pHe$ collisions in the ISM contribute significantly to the local $\bar{p}$ flux in the observed energy range, providing a very good fit of currently available data, and WIMP annihilations can be in principle a copious source of $\bar{p}$, antiprotons are a powerful channel to set limits on WIMP DM models. Still, as we just discussed, the prediction for the WIMP signal is severely affected by uncertainties in the propagation model and the DM distribution in the Galaxy.
In the following, taking the conventional astrophysical contribution (background) as obtained in the five propagation models listed in Table~\ref{tab:models} (see also Fig.~\ref{fig:bcmodels}-\ref{fig:test_source}), we consider the three DM WIMP scenarios introduced in Section~\ref{sec:dm} and derive constraints on the DM annihilation cross section, for a specific DM mass and our three reference spherical dark matter profile, by requiring that the total antiproton flux is within 3$\sigma$ to the combination of all the $\bar{p}$ flux data points. 

We clarify that those constraints are \textit{not} the most conservative constraints. In fact they are the strongest constraints we could get, by having propagation models that fit already the $B/C$ flux ratio, the $p$ and He fluxes and also give good fit to the $\bar{p}$ flux. Significantly weaker constraints on DM have been drown by allowing for greater uncertainties in $\bar{p}$ background flux \cite{Cirelli:2008pk, Donato:2008jk, Cholis:2010xb}. The most conservative upper limits on DM models come from being completely agnostic about $\bar{p}$ background fluxes, setting limits by demanding that the DM $\bar{p}$ flux does not exceed the measured $\bar{p}$ flux at any energy by more than 3$\sigma$ \cite{Cholis:2010xb}. Such a method provides more robust constraints. On the other hand the advantage of our method is that it provides \textit{more realistic} constraints. 

\begin{figure}[tbp]
\centering
\includegraphics[width=0.45\textwidth]{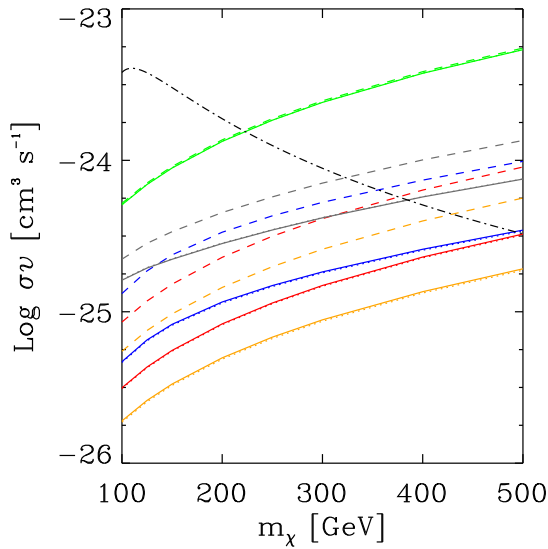}
\caption{Constraints for the Wino model as function of the particle mass. The black line corresponds 
to the cross section given in Eq.~\ref{eq:svwino}. Colors are as in Fig.~\ref{fig:bcmodels} (solid: Einasto 
profile, dotted: NFW, dashed: Burkert).}
\label{fig:constraints_wino}
\end{figure}

\begin{figure}[tbp]
\centering
\includegraphics[width=0.45\textwidth]{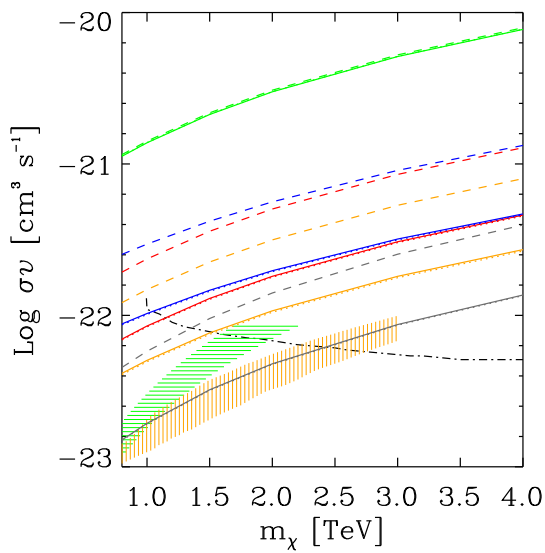}
\caption{Constraints for the heavy DM candidate in $\mu\mu$ channel. Colors are as in Fig.~\ref{fig:bcmodels} (solid: Einasto profile, dotted: NFW, dashed: Burkert). The orange shaded region is the Fermi $e^{+}+e^{-}$ data 3$\sigma$ fit region, and the green shaded region is the PAMELA positron fraction 3$\sigma$ fit region\cite{Bergstrom:2009fa}. The black line gives the HESS 2$\sigma$ upper limits \cite{Zaharijas:2010ca}.}
\label{fig:constraints_heavy_ww}
\end{figure}

\begin{figure}[tbp]
\centering
\includegraphics[width=0.45\textwidth]{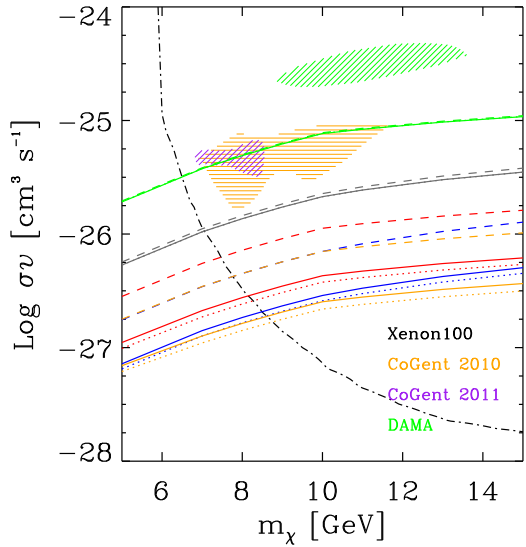}
\caption{Constraints for the light DM candidate in $b\bar{b}$ channel. Colors are as in Fig.~\ref{fig:bcmodels} (solid: Einasto profile, dotted: NFW, dashed: Burkert). Also shown for comparison the favored regions of annihilation cross sections connected to the 90$\%$ C.L. favored spin-independent 
elastic scattering cross sections regions from CoGent~\cite{Aalseth:2010vx, Aalseth:2011wp}, DAMA/LIBRA~\cite{Bernabei:2008yi} and the recent 90$\%$ C.L. limit from Xenon100~\cite{Aprile:2011hi}.}
\label{fig:constraints_bino}
\end{figure}

In Fig.~\ref{fig:constraints_wino} we present our 3$\sigma$ limits with three different spherical halo profiles (Einasto, NFW, Burkert), for the nonthermal Wino DM models up to 500GeV. 
The most tight constraints come from the thick (THK) propagation model, which probes a larger part of the dark halo, while the thin halo, for the opposite reason gives the weakest constraints similarly to the work by \cite{Hooper:2009fj}.
Yet even the thin diffusion model excludes a Wino DM lighter than 300 (200) GeV at 3$\sigma$ level for a Burkert (NFW) profile. Thus models such as \cite{Moroi:1999zb, Acharya:2008zi}, that have been suggested by \cite{Kane:2009if} to explain the rise of the positron fraction measured by PAMELA \cite{Adriani:2008zr} are excluded.
Note that the more conventional diffusion zone KRA and KOL models exclude Wino DM up to 500~GeV.   
In Fig.~\ref{fig:constraints_heavy_ww}, we give the equivalent constraints for heavy WIMPs that annihilate into $\mu^{+}\mu^{-}$ with the high energy muons that are produced emitting EW gauge bosons which are responsible for the antiproton yield \cite{Ciafaloni:2010ti}. While being an important source, the emission of the gauge bosons is not strong enough though, to exclude in most cases the regions of parameter space compatible at 3$\sigma$ with the fit of the PAMELA positron fraction and Fermi all-electron measurement~\cite{Zaharijas:2010ca}.
An interesting exception is the model with high convection, which excludes to 3$\sigma$ most part of the PAMELA 3$\sigma$ fit region above $m_{\chi}=1$ TeV.
Since the presence of convection, hardens the $\bar{p}$ fluxes, higher convection models can draw tighter constraints on the heavier DM models than low (or no) convection models do. 
This can clearly be seen by comparing the 3$\sigma$ limits from the convection model between Fig.~\ref{fig:constraints_wino},~\ref{fig:constraints_heavy_ww} and~\ref{fig:constraints_bino}. 
Thus to constrain leptophilic heavy DM models, via $\bar{p}$, we need to quantify better the level of convection in the Galaxy.

The updated upper limits from ARGO-YBJ \cite{DiSciascio:2011ic} (see also \cite{DiSciascio:2009ir}) at 2 TeV ($\bar{p}/p \leq 0.05$) and 5 TeV ($\bar{p}/p \leq 0.06$) do not put any tighter constraints on these heavy WIMPs either.
In Fig.~\ref{fig:constraints_bino} the results for a light WIMP annihilating to $b \bar{b}$ (to model for the cases of strong couplings to quarks) are presented. Also we show for comparison the favored/excluded regions of annihilation cross sections connected to the favored/excluded spin-independent elastic scattering cross sections through Eq.~\ref{eq:Yukcoupl}. The couplings of the DM scalar $\phi$ to the quarks $c_{q}$ -by contact interaction terms- are proportional to the Yukawa couplings. We show the equivalent region to the 90$\%$ C.L. favored region by CoGent in the data released in 2010~\cite{Aalseth:2010vx} and their more recent 2011 results~\cite{Aalseth:2011wp}, as well as the region favored by DAMA/LIBRA~\cite{Bernabei:2008yi} (without channeling). Independent studies have also analyzed the region favored by the CoGent dataset where an hint of annual modulation effect has been found, see, e.g., \cite{Schwetz:2011xm,Farina:2011pw,Fox:2011px,Hooper:2011hd}.
For instance, the results of Ref.~\cite{Hooper:2011hd} suggest a slightly higher WIMP-nuclear scattering cross section, which would also give in a slightly higher annihilation cross section; in Fig.~\ref{fig:constraints_bino} we present only the lower overall region related to~\cite{Aalseth:2011wp}. 
Finally we give the equivalent to the recent limit 90$\%$ C.L. from Xenon100~\cite{Aprile:2011hi}.
Our limits provide complementary constraints to direct detection limits below masses of 7 GeV. We note that, like Xenon100, our limits from all the models apart from the THN (thin halo) exclude the favored regions by CoGent and DAMA, while the THN model excludes only the DAMA region. This result  is similar (but more constraining) to the result in~\cite{Keung:2010tu} For a case where the DM particle is a vector, having also couplings to the Yukawa the CoGent and DAMA regions move down by a factor of 3, which are still strongly constrained by the data (for another analysis cross correlating antiproton and direct detection data for light WIMPs, see also~\cite{Lavalle:2010yw,Delahaye:2011,Cerdeno:2012}).
     
Also recently, \cite{Feng:2011vu, Frandsen:2011ts, DelNobile:2011je} have suggested the possibility of reconciling the CoGent and DAMA favored regions with the limits from CDMS and Xenon by having the coupling of DM to the proton vs the neutron different. 
This is done either from violation of isospin \cite{Feng:2011vu, Frandsen:2011ts}, or by having scatterings via both photon and Higgs echange \cite{DelNobile:2011je}.
These works suggest that the preferred by the data, value for the ratio of the effective coupling of the DM particle to the neutron $f_{n}$, to the proton $f_{p}$, is $f_{n}/f_{p} \sim -0.7$ ($-0.71$ for \cite{DelNobile:2011je}). 
Yet since in all these models, the suggested scattering cross section to the proton (that agrees with all the data) is higher by about 2 orders of magnitude compared to that for a scalar DM with $f_{n}/f_{p}=1$ as shown in Fig.~\ref{fig:constraints_bino}, these models are strongly disfavored by the $\bar{p}$ constraints.

\section{More on astrophysical uncertainties}\label{sec:exotic}

The constraints shown in section \ref{sec:constraints} already give a clear evidence of the relevance of the associated uncertainties. On the one hand, predictions of DM antiprotons suffer from uncertainties due to the unknown density distribution of the DM toward the galactic center. On the other hand, the effective propagation models determined by fitting the nuclear CR components lead to different predictions for DM originated and, to less extent, astrophysics generated antiprotons. 

Moreover, we should remark that even a very precise determination of the local effective propagation parameters in Table~\ref{tab:models} would leave large uncertainties 
on the propagation conditions in the inner Galaxy, where the DM production rate is also maximal (unlike the standard astrophysical $\bar{p}$). Therefore, the predicted $\bar{p}$ flux from DM is overall more uncertain than the astrophysical $\bar{p}$ flux, or, for that matter, any CR spectrum from the conventional astrophysical sources. Yet we remark that predictions of electron and positron spectra from dark matter are instead less sensitive to these uncertainties, because the electron/positron mean free path at high energy is shortened by strong energy losses.

In order to better discuss these points, we show in the following the resulting effect of either modifying the CR propagation properties in the inner Galaxy or introducing a disklike structure in the DM density distribution. 

\subsection{Nonstandard propagation models}\label{sec:nsprop}

We propose here a few \emph{nonstandard} diffusion models to show to what extent we can change the physical conditions within the inner $3~\kpc$ of the Galaxy to modify the fluxes from DM without affecting significantly the standard observables. Effects of different position dependence for the diffusion coefficient or of a different profile for the convection velocity were already investigated in \cite{Grajek:2010bz} and in \cite{Perelstein:2010gt}. It is worth emphasizing here that this analysis can be performed only with numerical models, and in particular exploiting the features of the \dragon\ code, as opposed to the semianalytic approach within which it is not possible to account for a variation of the diffusion coefficient and of the convective velocity with position.

For reference, our exotic propagation models are based on the KRA model, suitably modified as described below:
\begin{itemize}
\item in the \emph{expr} model the diffusion coefficient is assumed to be very small in the inner galaxy $R<3~\kpc$, according to  
\begin{equation}
D(r) = D_{0}\times0.5\times\left[\tanh\left((r-3~\kpc)/0.25~\kpc\right)+1.02\right]
\label{eq:expr}
\end{equation}
Local variations of the diffusion coefficient are naturally expected, due to the position dependence of the magnetic turbulence injection and transport. While to explain nuclear CR data such radial dependence needs not be invoked, the explanation of nonlocal observables, as for example the $\gamma$-ray profile the gradient problem found in EGRET and Fermi~\cite{Collaboration:2010cm} data, may be rather natural in models with radial variation of the diffusion coefficient \cite{Evoli:2008dv}.

\item in the \emph{convective} model we assume instead that a strong convective wind is effective in the inner region $R<3~\kpc$. The convective wind is assumed to be directed outside the galactic plane in the $z$-direction, with an intensity $v_C(z) = 200\times (z/1~\kpc)~\km/\s$. We remark that according to ROSAT observations such strong stellar winds can exist in the inner galaxy and affect CR propagation \cite{Gebauer:2009hk}. In the same paper \cite{Gebauer:2009hk} such winds were also proposed as a possible alternative solution to the ``CR gradient problem''.
\end{itemize}
Besides these two models we also tested a {\em Gaussian bubble} configuration, with the diffusion coefficient having a Gaussian increase away from the solar system as
\begin{equation}
D(r,z) =  D_{0}\, e^{\left(\frac{r-R_{\odot}}{r_{D}}\right)^{2}}e^{\left(\frac{\mid z \mid}{z_{D}}\right)^{2}}.
\label{eq:gauss_tor}
\end{equation}
We considered the cases $r_D \leq 5~\kpc$ and $z_D \leq 5~\kpc$, thus making the escape time of locally produced CRs much larger than that of CR produced at $|z| > z_{D}$, $r-R_{\odot} > r_{D}$. As a result the observed $p, \bar{p}$  spectra at high energies are dominated by local sources while at lower energies more distant sources are more dominant even for CR $p$ and $\bar{p}$. However, while by selecting properly the injection spectra for the protons we can recover the observed proton spectrum, the spectrum of the B/C ratio can not be recovered by changing the diffusion parameters, Alfv\'en speed  and nuclei injection spectrum properties within reasonable values. Thus diffusion models such as that of Eq.~\ref{eq:gauss_tor} fail in fitting secondary fluxes overall and we will not discuss them any longer.

\begin{figure}[tbp]
\centering
\includegraphics[width=0.45\textwidth]{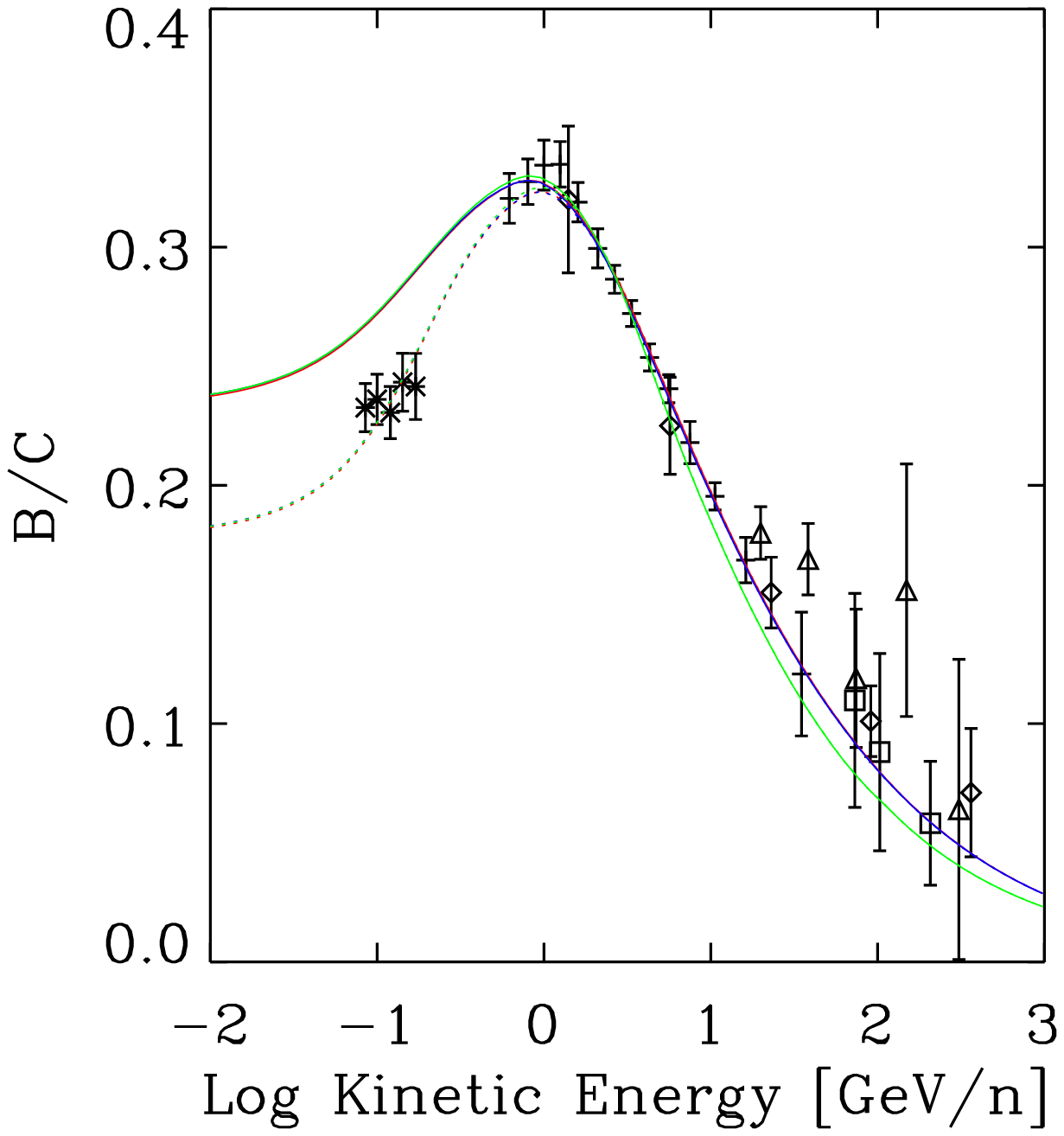}
\includegraphics[width=0.45\textwidth]{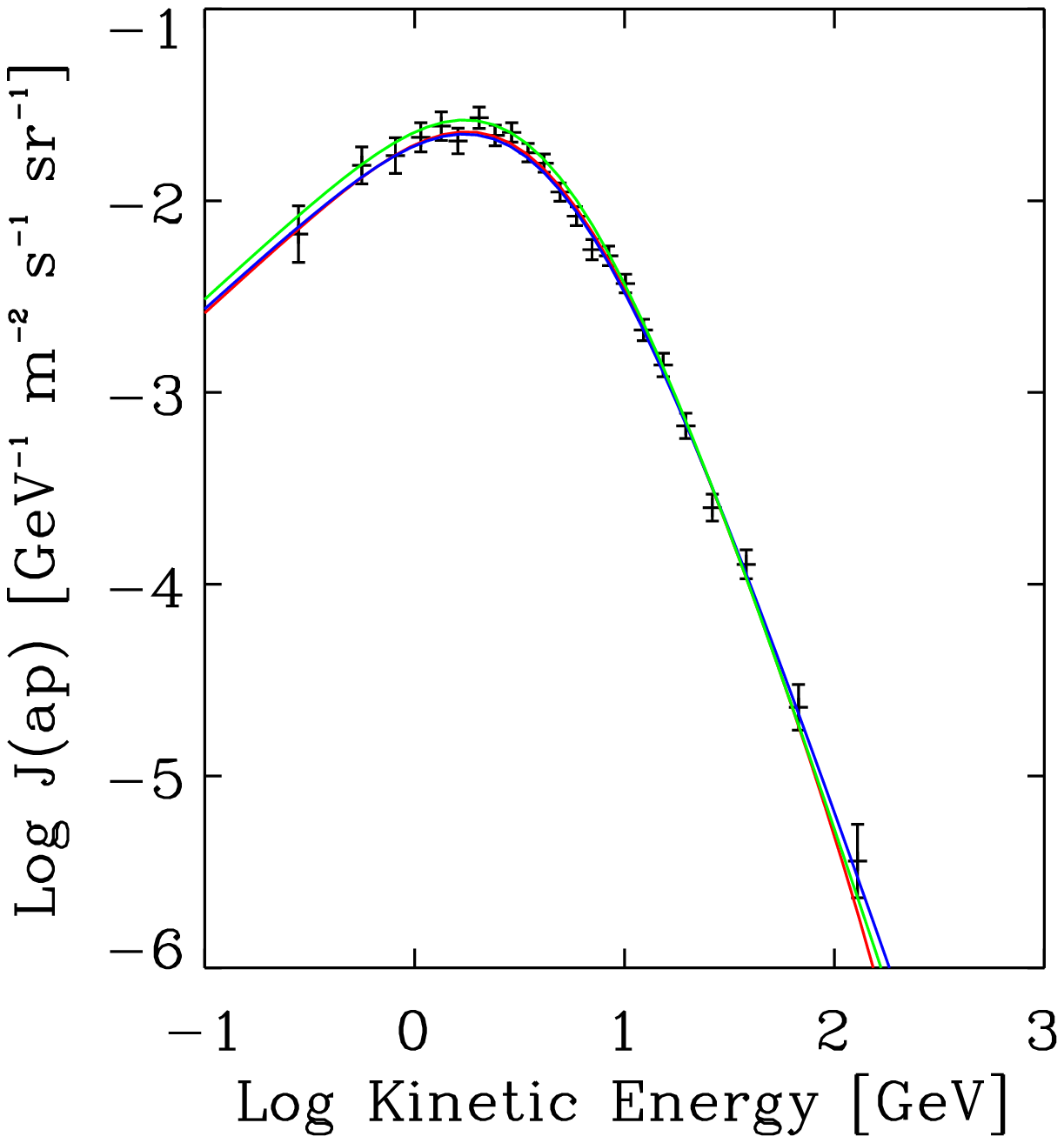}
\caption{{\it Left panel}: The B/C for the KRA model (red), for the model with radial dependent diffusion coefficient (green) and the model with strong convection in the inner galaxy (blue). We remark that the red and blue lines are superimposed. See Sec.~\ref{sec:nsprop} for details. {\it Right panel}:  the secondary $\bar{p}$ flux computed for the same models.}
\label{fig:exoticbc}
\end{figure}

\begin{figure}[tbp]
\centering
\includegraphics[width=0.9\textwidth]{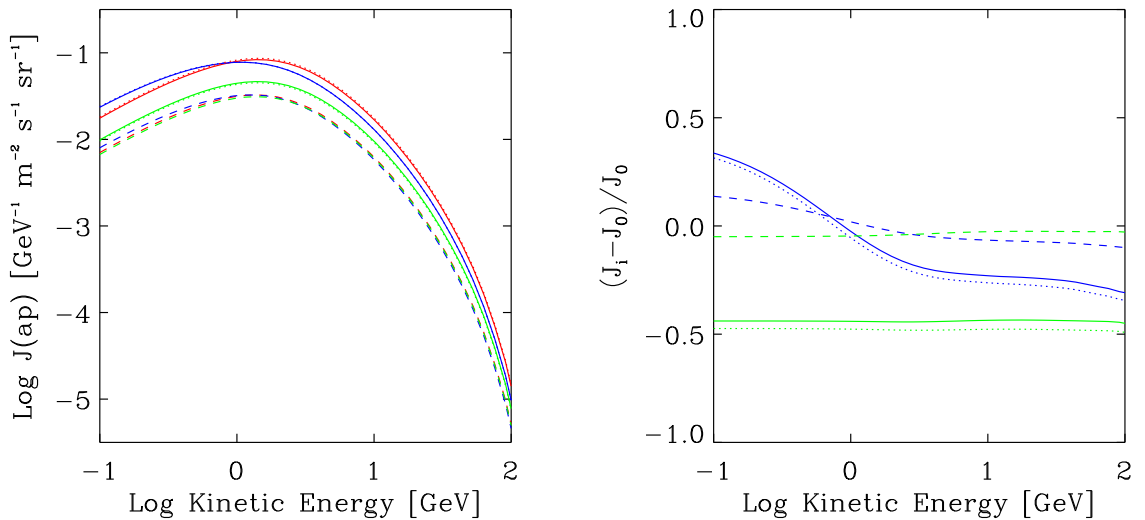}
\caption{{\it Left panel}: The $\bar{p}$ flux from a DM Wino of 200 GeV for the KRA model (red), for the model with radial dependent diffusion coefficient (green) and the model with strong convection in the inner galaxy (blue), for different DM profiles (solid: Einasto, dotted: NFW, dashed: Burkert). {\it Right panel}: The relative ratio of the nonstandard propagation models with the KRA model (same color/style code as for the left panel).}
\label{fig:exoticap}
\end{figure}

Results are shown instead for the {\em expr} and the {\em convective} models in Fig.~\ref{fig:exoticbc} for B/C and astrophysical antiprotons and in Fig.~\ref{fig:exoticap} for DM antiprotons. As it is clear from Fig.~\ref{fig:exoticbc} the B/C ratio and the secondary antiproton spectrum are very little affected by the propagation conditions within the inner galaxy. This can be understood by considering that the typical interaction timescale for B, C and ordinary protons (which then originate the secondary antiprotons) is of order $\tau \sim (n_g\, c \,\sigma)^{-1} \sim 6\times10^{14}~\s ~(n_g/1~\cm^{-3})^{-1}(\sigma/50~{\rm mb})^{-1}$, which yields a typical propagation length of order $\lambda(\rho) \sim \sqrt{6\,D(\rho)\,\tau} \sim 5.4~\kpc ~(\rho/20~{\rm GV})^{0.25}$ in the KRA model. Moreover, astrophysical B, C and $\bar{p}$ sources are mainly correlated with SNR and gas distributions, which peak at $R\simeq 4~\kpc$. Therefore significant contribution of B and C at the Earth position cannot come from the inner galaxy at energies below a few 10 GeV/n. At energies above a few 10 GeV/n the propagation length is larger and contribution from the inner region becomes more relevant, although the B/C ratio is still little dependent upon physics in that region. The same argument applies to astrophysically generated antiprotons.

On the other hand, the DM $\bar{p}$ source distribution clearly peaks at the GC. Therefore, the DM $\bar{p}$ flux is indeed strongly sensitive to the propagation conditions in the GC region, as it is clear from Fig.~\ref{fig:exoticap}. In particular, for spiked profiles the effect can be as large as $\sim80\%$ at 100 GeV, while for cored profiles the effect is much smaller and probably comparable with other uncertainties (see right panel in Fig.~\ref{fig:exoticap}). 
\subsection{Effects of a possible dark disk}
The presence of a possible disk DM structure (DD) has been recently suggested by \cite{Read08} as a natural expectation in $\Lambda$CDM scenarios. We therefore considered also this possibility. When propagating CR $\bar{p}$ accounting for the presence of a dark disk, the DM density is assumed to be $\rho = \rho_{SH} + \rho_{DD}$, with $\rho_{SH}$ the DM density of the 
spherical halo. We recall that since we are interested in understanding what is the maximal effect that the dark disk can have on the $\bar{p}$ flux, we set for simplicity $\rho_{0_{DD}}/\rho_{0_{SH}} = 1$ in our simulations. We also set the total local DM density to be $0.4~\GeV \, \cm^{-3}$, as we have done in the case of a spherical only component. 

As we show in Fig.~\ref{fig:test_source} (bottom row), the $\bar{p}$ flux of DM origin produced at distances larger than 6 kpc from our position can be rather important. In fact assuming a NFW profile we get that $\simeq 40 \%$ of the observed DM $\bar{p}$ flux is produced from the inner 2 kpc of the halo,\footnote{With the exception of the thin halo (THN) model.} where including a dark disk component with $\rho_{0_{\rm DD}} = \rho_{0_{\rm SH}}$ has the biggest effect on the density.
\begin{figure}[tbp]
\centering
\includegraphics[width=0.9\textwidth]{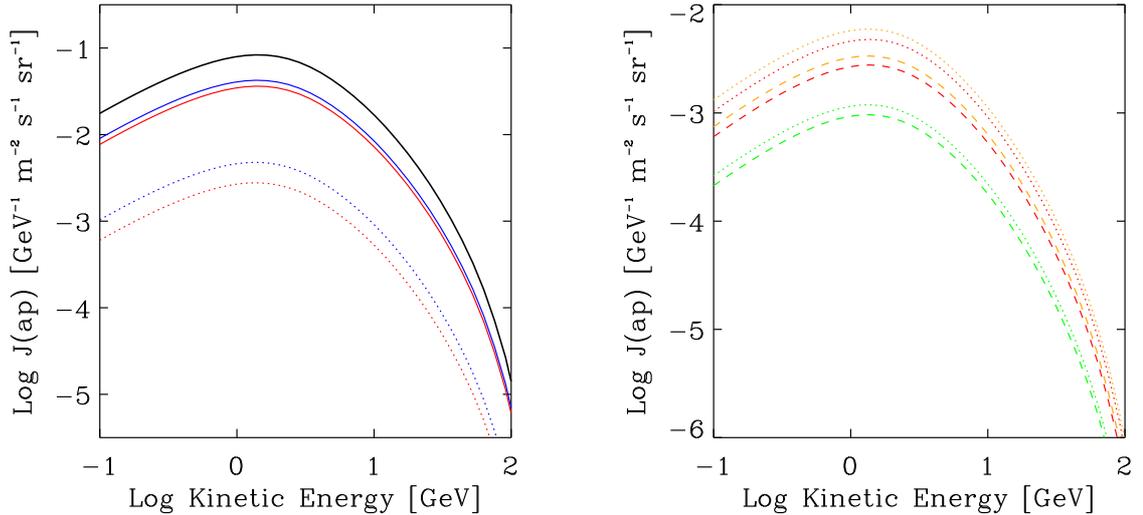}
\caption{{\it Left panel}: The $\bar{p}$ flux from a 200 GeV Wino obtained for the KRA model assuming different DD profiles [red lines are for the model of Eq.~\ref{eq:DD1}, blue lines are for the model of Eq.~\ref{eq:DD2}]. The contribution from DD only is shown as dotted lines, the total contribution is shown as solid lines. For reference, the black solid line shows the $\bar{p}$ for the NFW profile without a DD profile. 
{\it Right panel}: The $\bar{p}$ obtained for the propagation model with different $z_{t}$: KRA (red), THK (yellow) and THN (green). Dashed line is for the model of Eq.~\ref{eq:DD1}, dotted for the model of Eq.~\ref{eq:DD2}.}
\label{fig:apDD}
\end{figure}
In Fig.~\ref{fig:apDD} (left), we show the $\bar{p}$ flux of DM origin for the case of combined $\rho_{\rm SH}+\rho_{\rm DD}$ profile, and compare it to the spherical only (NFW) profile. As a propagation model we use our KRA model which has $z_{t} = 4$~kpc, that is significantly larger than the $z_{H}$ scales of the DD profiles. The resulting DM originated $\bar{p}$ flux is decreased by a factor of 2 in both considered cases, due to the fact that the difference between a spherical halo only source and a spherical halo + dark disk source is of that order in the relevant diffusion region.

Interestingly, the dark disk contribution ($\rho_{\rm DD}^{2}$) to the $\bar{p}$ flux is a factor of 10 times lower than that of the spherical component in the entire range of the $\bar{p}$ spectrum (see dotted lines in Fig.~\ref{fig:apDD} left). Such a suppression happens since, as we show in Fig.~\ref{fig:test_source} (bottom row) only $\simeq 10-20 \%$ of the DM $\bar{p}$ flux in the case of a spherical halo is produced within the inner 1 kpc from the Sun. With the rest of the $\bar{p}$ flux produced at distances $> 1$~kpc from the Sun's position, having a dark disk component with much lower DM density (than the spherical halo) especially outside the disk plane (but within the diffusion zone) can have a strong impact on the DM $\bar{p}$ flux (see also discussion in \cite{Cholis:2010px} for the case of Sommerfeld enhanced annihilating DM). We note that while the different dark disk profiles vary in their $\rho_{\rm DD}^{2}$ component of the flux, the total $(\rho_{\rm SH} + \rho_{\rm DD})^{2}$ does not vary that much between those models, thus making the details of the dark disk profile less important (as long as we are in the regime of $z_{t} > z_{H}$).

In Fig.~\ref{fig:apDD} (right), we show the effect of different propagation models in the \textit{only} $\rho_{DD}^{2}$ component of the flux, for the two considered dark disk profiles. Depending on the DD profile, the effect of changing the propagation model (for a fixed DD profile) is between a factor of $\simeq 6$ [model of Eq.~\ref{eq:DD1}] and a factor of $\simeq 4$ [model of Eq.~\ref{eq:DD2}]. Adding the $\rho_{SH}^{2}$ and $2 \rho_{SH} \rho_{DD}$ terms
of the flux makes though the changes in the total DM $\bar{p}$ flux between different models similar to those shown in Fig.~\ref{fig:apdm} (with the fluxes though suppressed by a factor of 2).
 
\section[]{Discussion: comparison with previous results and future perspectives} \label{sec:discuss}

\subsection{Comparison with the semianalytic solution}

Most analyses of the antiproton DM signal in the literature have been performed via semianalytic solutions of the diffusion equation. Equation~\ref{eq:diffusion_equation} admits such solutions in the case of a set of simplifying assumptions is implemented. On top of the hypotheses of stationary limit, cylindrical symmetry and free escape at the boundary of the diffusion region, which are applied also to our numerical treatment, one needs to restrict to models with: {\sl (i)} a spatially constant diffusion coefficient (as opposed to exponentially rising in the vertical direction, and, possibly, with a radial dependence within our model); {\sl (ii)} an infinitely thin gas disc of constant density located at $z=0$ (rather, again, than some more realistic gas distribution given as a function of $R$ and $z$ -- a disc with finite thickness is also an option, and in this case the diffusion coefficient can be set to two different functions of rigidity, one in the disc and one in the diffusive halo); {\sl (iii)} energy losses and reacceleration which are either neglected or included as a matching solution term at $z=0$; and  {\sl (iv)} a convection velocity perpendicular to the Galactic plane and with the specific form $\bm{v}_{c} = (0,0, sign(z)\,V_c)$, where $V_c$ is a constant (however, we will not compare convection models in the following). Under these hypotheses, the transport equation can be worked out by factorizing out the radial part of the solution through an expansion of the number density and its source function in a Fourier-Bessel series, and solving analytically the differential equation in the vertical direction on each term of the series. The solution can then be written as a sum of Fourier-Bessel modes, with integrals over volume of the source function entering in the coefficients. This is the solution implemented, e.g., in the \ds\ package to estimate the local antiproton flux from WIMP annihilations. If the DM density profile is not too large in the GC region, the sum converges rapidly and this approach to the diffusion equation can be substantially less CPU consuming than the full numerical solution. For very cuspy distributions, the terms in the series show an oscillatory pattern which instead converges rather slowly and this semianalytic approach may not be anymore that powerful (see, however, the procedure to improve the convergence suggested in~\cite{BringmannThesis}).
\begin{figure}[tbp]
\centering
\includegraphics[width=0.9\textwidth]{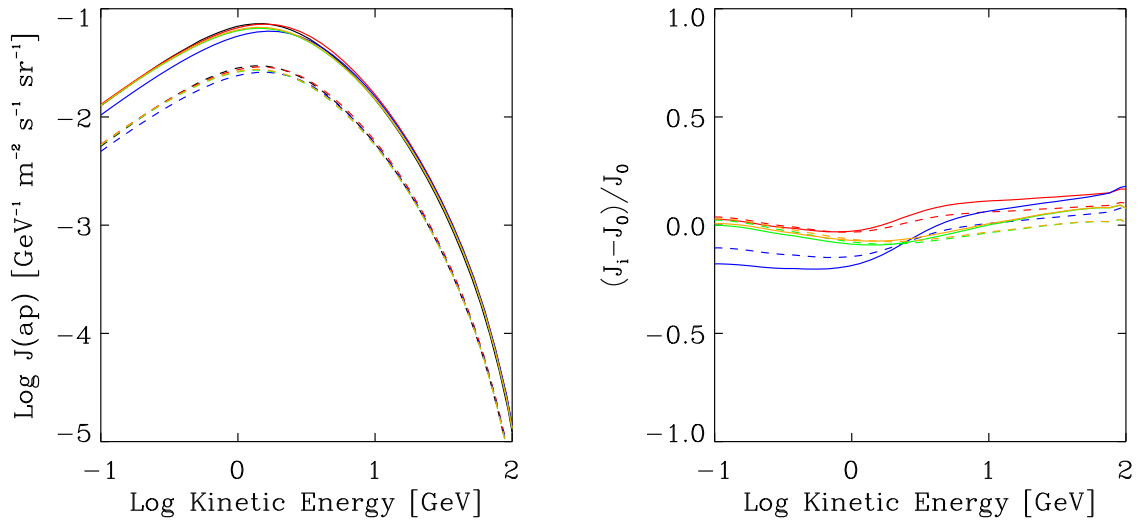}
\caption{{\sl Left panel:} Local spectrum of antiprotons from 200 GeV Wino DM for the KRA model and assuming a modulation potential of 550 MV, obtained by: running the semianalytic model in \ds\ (black); running \dragon\ in our standard setup (red); with a \dragon\ run which assumes a spatially constant diffusion coefficient (blue), plus no reacceleration (green), plus no energy losses (orange). Solid lines are obtained assuming Einasto profile and dashed lines are for Burkert profile. {\sl Right panel:} Fraction ratio between the \dragon\ runs (same color coding) and the \ds\ result.}
\label{fig:analytic1}
\end{figure}
In Fig.~\ref{fig:analytic1} we show, for the 200~GeV Wino model already discussed in Secs.~\ref{sec:wino} and \ref{sec:local} and the spherical Burkert and Einasto profiles, a comparison between our numerical solution in the case of the KRA propagation model and the solution obtained with the \ds\ code trying to match as closely as possible the propagation setup. In particular, we assume that the height of the diffusion region corresponds to our parameter $z_{t}$~\cite{DiBernardo10}, a constant diffusion coefficient, namely setting $g(R,z)=1$ in Eq.~\ref{eq:diff_coeff} and keeping the same dependence on rigidity, that the gas (hydrogen) has a density of 1~cm$^{-3}$ in a layer of 0.2~kpc thickness (but implemented in the solution as infinitely thin disc at $z=0$), that energy losses and reacceleration can be neglected. In the left panel of Fig.~\ref{fig:analytic1} we plot the result of the computation obtained with \ds\ (black lines), the result with our standard \dragon\ run (red lines) and results with other \dragon\ runs as obtained changing progressively the configuration to a constant diffusion coefficient within $z_{t}$, switching off reacceleration and energy losses. As it can be seen, in this case the results of the computation in the semianalytic and the numerical model are in fairly good agreement, within about 20\%. One can also see that there is no conspiracy between the different simplifying assumptions to induce a larger or smaller antiproton flux, but they tend to compensate; in particular, the effect of neglecting reacceleration is rather mild as one could have expected from the broad shape of the WIMP source function, and energy losses have really no impact.
\begin{figure}[tbp]
\centering
\includegraphics[width=0.9\textwidth]{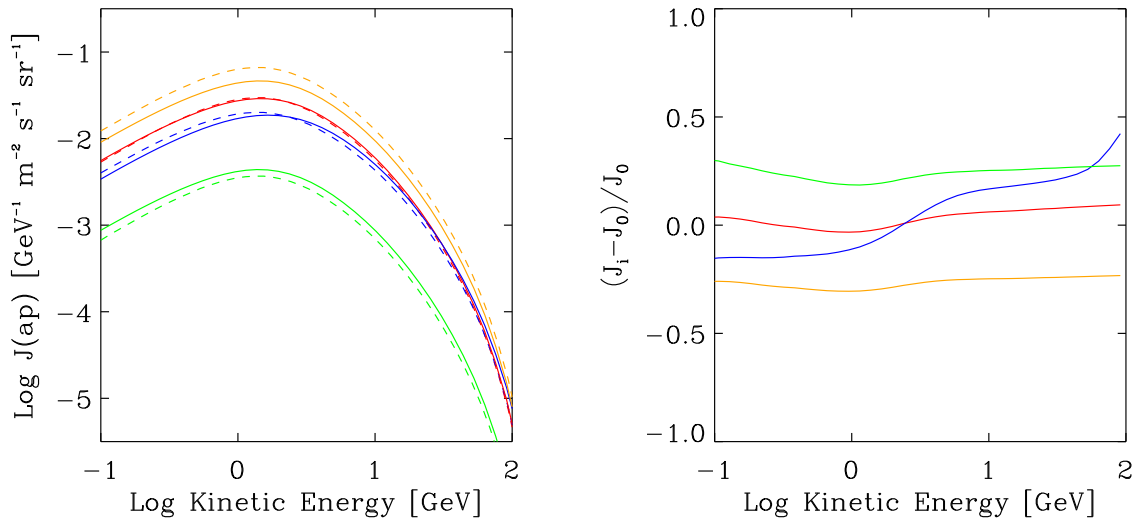}
\caption{{\sl Left panel:} Comparison between results obtained with the semianalytic propagation code included in \ds\ (dashed lines) and standard runs with \dragon\ (solid lines), for a 200 GeV Wino DM and for our standard set of propagation models (the color coding is the same as in Fig.~\ref{fig:bcmodels}). We show results for the Burkert profile. {\sl Right panel:} Fraction ratios between the \dragon\ runs and the \ds\ results.}
\label{fig:analytic2}
\end{figure}
In Fig.~\ref{fig:analytic2} we show, for the same WIMP model, the comparison of the full \dragon\ numerical solution with the \ds\ solution, for some of the propagation models in Table~\ref{tab:models} and the Burkert spherical halo profiles. Even for the Burkert profile, for which we had
found that the largest contribution to the locally measured flux comes from local sources, the prescription we have implemented for translating
one model into the other does not work as well as for the first example we have considered, with differences of the order of 40\%.

\subsection{Comparison with previous results}

The literature on the ${\bar p}$ based constraints on DM models is quite wide. Here we limit ourselves to compare our results with some of the most recent results. 
In \cite{Donato:2008jk} the PAMELA ${\bar p}/p$ data were first used in this framework to constrain WIMP annihilation cross section in a mass range between $\sim100~\GeV$ and 1 TeV. The DM ${\bar p}$s as well as the secondaries' propagation were treated with a semianalytical code as described in \cite{Donato01,Donato:2003}.  The uncertainty on the secondary flux was evaluated to be 20-30\%, which is comparable with our results, 
while that on the DM contribution due to propagation amounts to about 1 order of magnitude (see also \cite{Barrau:2005,Bringmann:2006im}), which is slightly smaller than that found in this paper. 
The constraints in \cite{Donato:2008jk} are based on a bin by bin comparison of theoretical models and experimental data which is different from the quality of fit analysis performed in this work. We found however that our constraints on the Wino models are in rough agreement with those results.  Heavy leptophilic and light WIMP models were not considered in  \cite{Donato:2008jk}.  Furthermore electroweak corrections to ${\bar p}$ production were not considered.

Light WIMP models were constrained in \cite{Bringmann:2009ca}. Again, propagation was treated semianalytically and electroweak corrections were not included in that analysis.  Our constraints are slightly more stringent than those in that paper for most propagation models but the THN one. 

In \cite{Cholis:2010xb} one of us used the \galprop\ numerical diffusion package to propagate secondary as well as DM ${\bar p}$s for a nonthermal wino model
with a propagation setup similar to our KOL. By comparing the sum of those two contributions with PAMELA data he obtained constraints very similar to those corresponding to the blue lines in Fig.~\ref{fig:constraints_wino}.  Electroweak corrections, which were not considered in that paper, have a marginal role for that class of models. Those corrections were instead included in the analysis reported in \cite{Garny:2011cj} and \cite{Cirelli:2010xx}, both based on semianalytical diffusion models.  The antiproton based constraints which were obtained considering the  $\chi \chi \rightarrow W e\nu$  and $\chi \chi \rightarrow Z e^- e^+$ annihilation channels (the ${\bar p}$s being produced by the gauge boson decay) are similar to those derived in this paper for the wino models.  Heavy {\it leptophilic} models (for which the effect of electroweak corrections are the largest), however, were not considered in  \cite{Garny:2011cj} and no constraints on the cross sections were derived in \cite{Garny:2011cj}.

\subsection{The projected AMS-02 sensitivity} 

As we discussed in Sec.~\ref{sec:cr_models} the present uncertainty on the propagation parameters gives rise to large scatter in the secondary and especially in the antiproton spectra from DM. This strongly limits the present sensitivity to dark matter indirect search in that channel.
Here we shortly discuss how the AMS-02 observatory, which was deployed on the International Space Station in May 2011, may drastically improve this situation.

AMS-02 is designed to provide simultaneous measurements of a number of CR species, including antiprotons, protons and a wide set of nuclei, with unprecedented precision.  In order to estimate the sensitivity of this observatory to DM models in the ${\bar p}$ channel we adopt the following preliminary AMS-02 performances\footnote{M.~Incagli,  B.~Bertucci, private communication.}.

We will limit ourselves to energies below 250 GeV in which range we use the antiproton geometrical acceptance  $A_{\bar p} = 0.25~\m^2 \sr$.  The energy resolution for protons and antiprotons is expected to be $\Delta E/E \sim 20~\%$ at about 100 GeV and to become $10~\%$ below 10 GeV. We conservatively assume $\Delta E/E \sim 25~\%$ below few hundred GeV which allows 10 bins per energy decade.  We also use a projected geometrical acceptance for light nuclear species $A_{\bar N} = 0.45~\m^2 \sr$ in order to estimate the AMS-02 sensitivity to CR propagation parameter.  We note that at the highest $\bar{p}$ energies the geometrical acceptance is expected to drop from the value of 0.25~${\rm m}^2\,{\rm sr}$, while the energy resolution become worse. Most importantly proton spill-over from soft scatterings inside the detector is expected to increase the observed $\bar{p}$ fluxes, thus place a limit to the highest energies at which a reliable measurement of the $\bar{p}$ flux can be made.  

As we discussed in Sec.~\ref{sec:cr_models}  the present uncertainty on the propagation parameters give rise to a large scatter in the secondary and especially in the antiproton from DM annihilation computed spectra. Those uncertainties are expected to be considerably reduced by AMS-02 thanks to its planned accuracy measuring several secondary/primary nuclear species ratios,  the B/C most importantly. In Ref.~\cite{Pato:2010ih} the AMS-02 projected error on the measurement of the diffusion coefficient spectral index was estimated to be  $\Delta \delta \simeq 0.02$ and that on the halo scale height $\Delta z_t =  1~\kpc$ with the same confidence.\footnote{Although a vertically homogeneous diffusion coefficient was adopted in that paper those errors should not change significantly adopting a exponential vertical profile. We also note that these results where calculated for fixed ISM gas, convection and re-acceleration assumptions. Relaxing them may have a more strong impact on how much $\Delta \delta$ and $\Delta z_t$ will decrease by the AMS-02.} Although in that paper the AMS-02 superconductor design was adopted,  we estimated that the uncertainty on $\delta$ should not change significantly with the final AMS-02 design. However, in order to account for possible systematics, we adopt here the larger error  $\Delta \delta \simeq 0.05$.
  
Even under these conservative assumptions, the error with which AMS-02 should be able to determine $\delta$ will be considerably smaller than the present uncertainty ($ 0.3 \lesssim \delta \lesssim 0.7$ \cite{DiBernardo10}) allowing to drastically reduce the allowed set of propagation models.

The uncertainty with which AMS-02 will constrain $z_t$ may be larger than that estimated in \cite{Pato:2010ih} as a consequence of the reduced mass resolution of the new experimental design.  This turns into a smaller discriminating power nuclear isotopes in particular of the $^{10}$Be and $^{9}$Be the  ratio of which can used to infer the diffusive halo thickness.   Since the AMS-02 collaboration has not released yet the mass resolution of the new instrumental setup, here we consider several test values of $\Delta z_t$.  

\begin{figure}[tbp]
\centering
\includegraphics[width=0.45\textwidth]{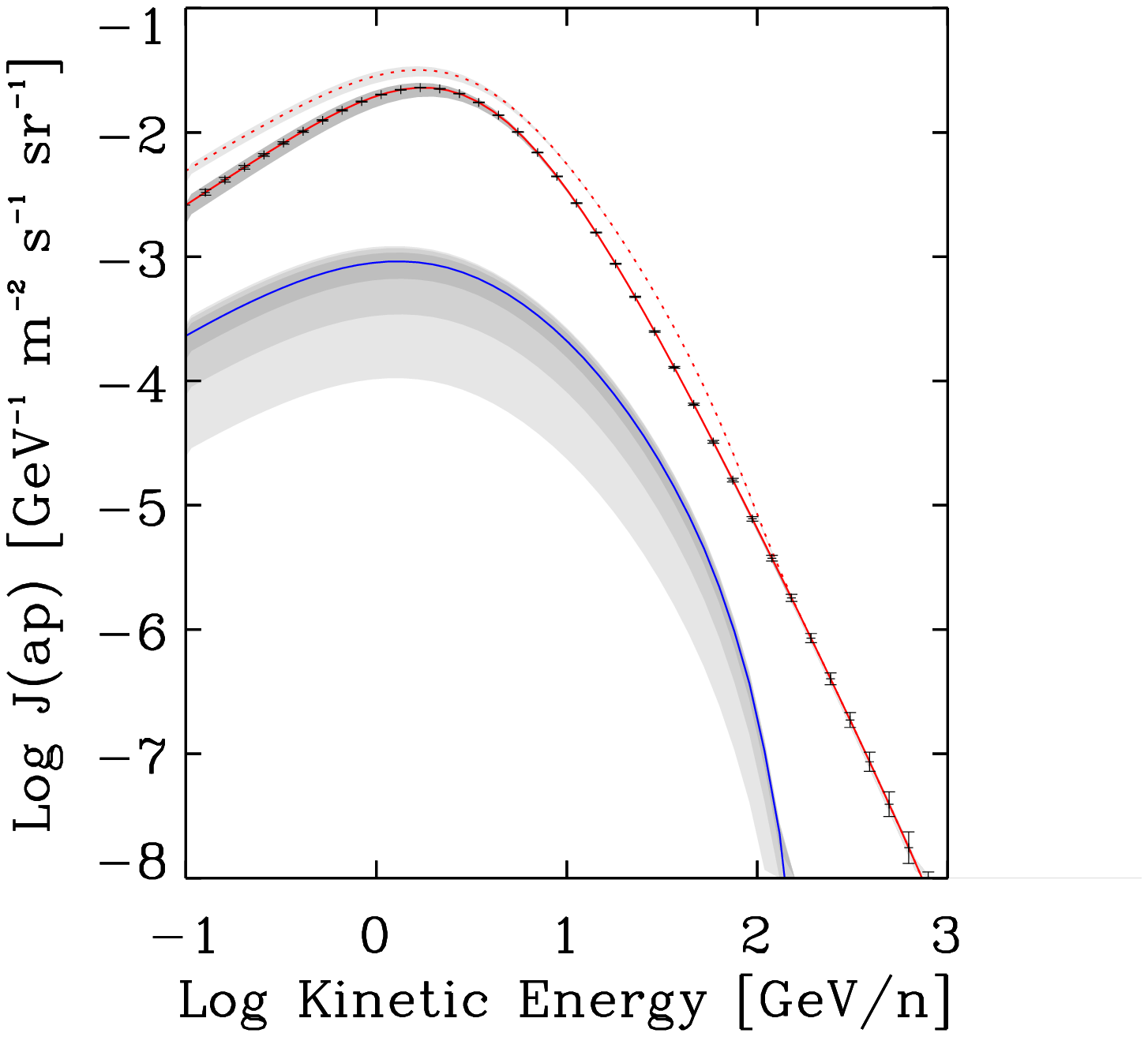}
\caption{The ${\bar p}$ spectrum from Wino DM with $m_\chi = 200~\GeV$ and annihilation cross section $\sv = 3 \times 10^{-25}~\cm^3 \s^{-1}$ propagated according to the KRA model (blue solid line); the secondary ${\bar p}$ computed for the same model (red solid line); the sum of those components (red dotted line). Error bars represent AMS-02 projected statistical errors after 1 year of data taking. The gray band around the red solid line is the uncertainty on the secondary ${\bar p}$ spectrum due to the expected AMS-02 error on the B/C measurement. The light, medium and dark grey bands around the blue solid line represent the uncertainties corresponding to an error $\Delta z_t = 1, 2, 3~\kpc$ on the knowledge of the diffusive halo scale height.}
\label{fig:ams_spectra}
\end{figure}

For illustrative purposes we assume that the actual propagation setup is described by the KRA model.  Under this hypothesis, in Fig. \ref{fig:ams_spectra} we compare the total (secondary + DM) ${\bar p}$ spectrum computed for a Wino DM models with $m_\chi = 200~\GeV$ and a cross section $\sv  = 3 \times 10^{25}~\cm^3\s^{-1}$, with the secondary ${\bar p}$ spectrum. The AMS-02 projected error refer to 1 year of data taking. 
The band about the predicted secondary spectrum (red dashed line) represents the projected uncertainty corresponding to varying $\delta$ in the range $0.5 \pm 0.05$ while tuning the other parameters so to keep the B/C compatible with the model within $2~\sigma$. We see that this uncertainty, which is mainly due to degeneracy of the B/C ratio for different values of the Alfv\'en velocity, is relevant only below few GeV. 

The three bands representing the ${\bar p}$ flux from DM, correspond to  $\Delta z_{\rm t} =  1, 2, 3~\kpc$. Since the effect of this uncertainty on the total ${\bar p}$ flux is very small, in the figure we show only that corresponding to $\Delta z_{\rm t} =  3~\kpc$.  

\begin{figure}[tbp]
\centering
\includegraphics[width=0.45\textwidth]{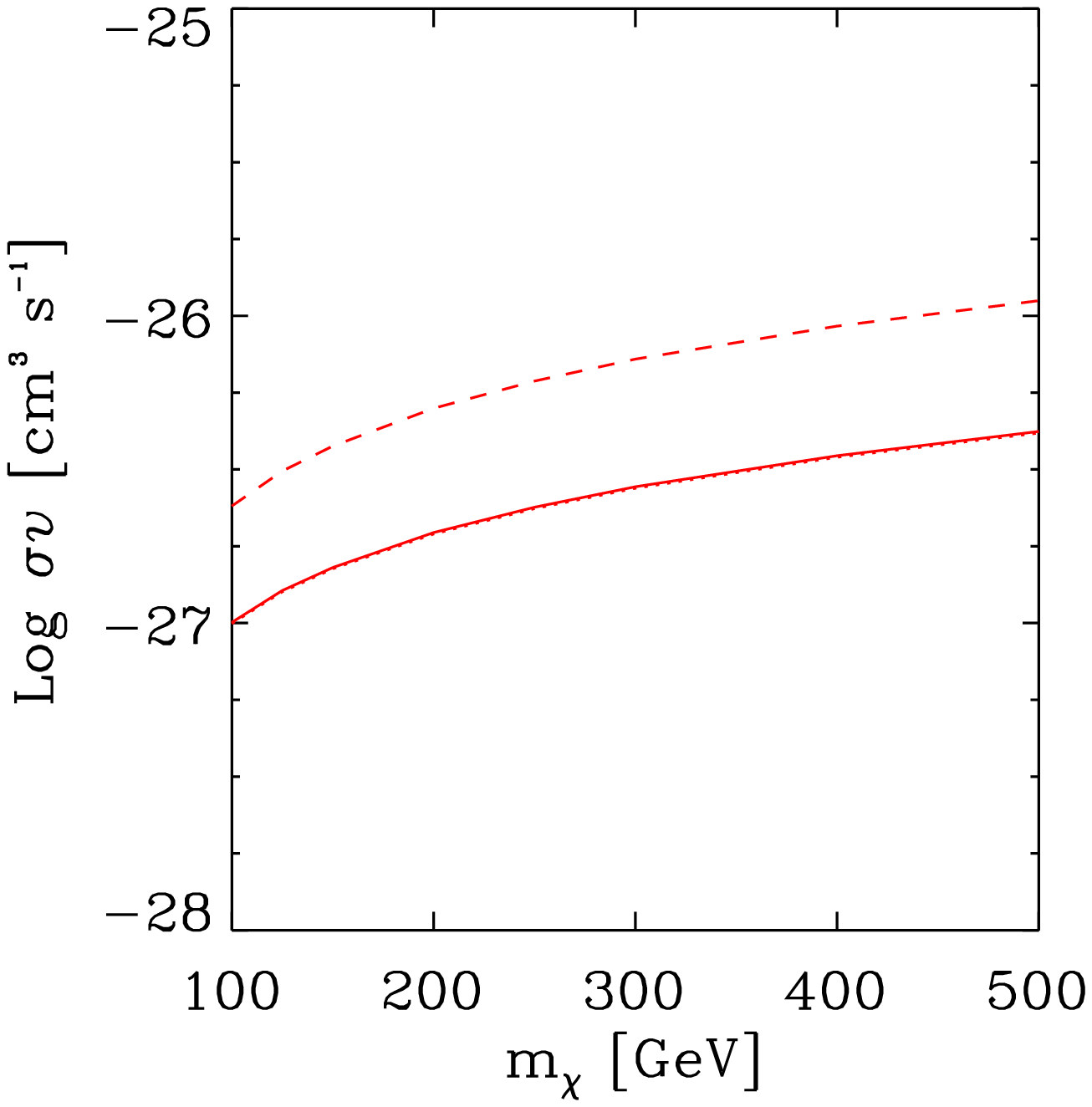}
\includegraphics[width=0.45\textwidth]{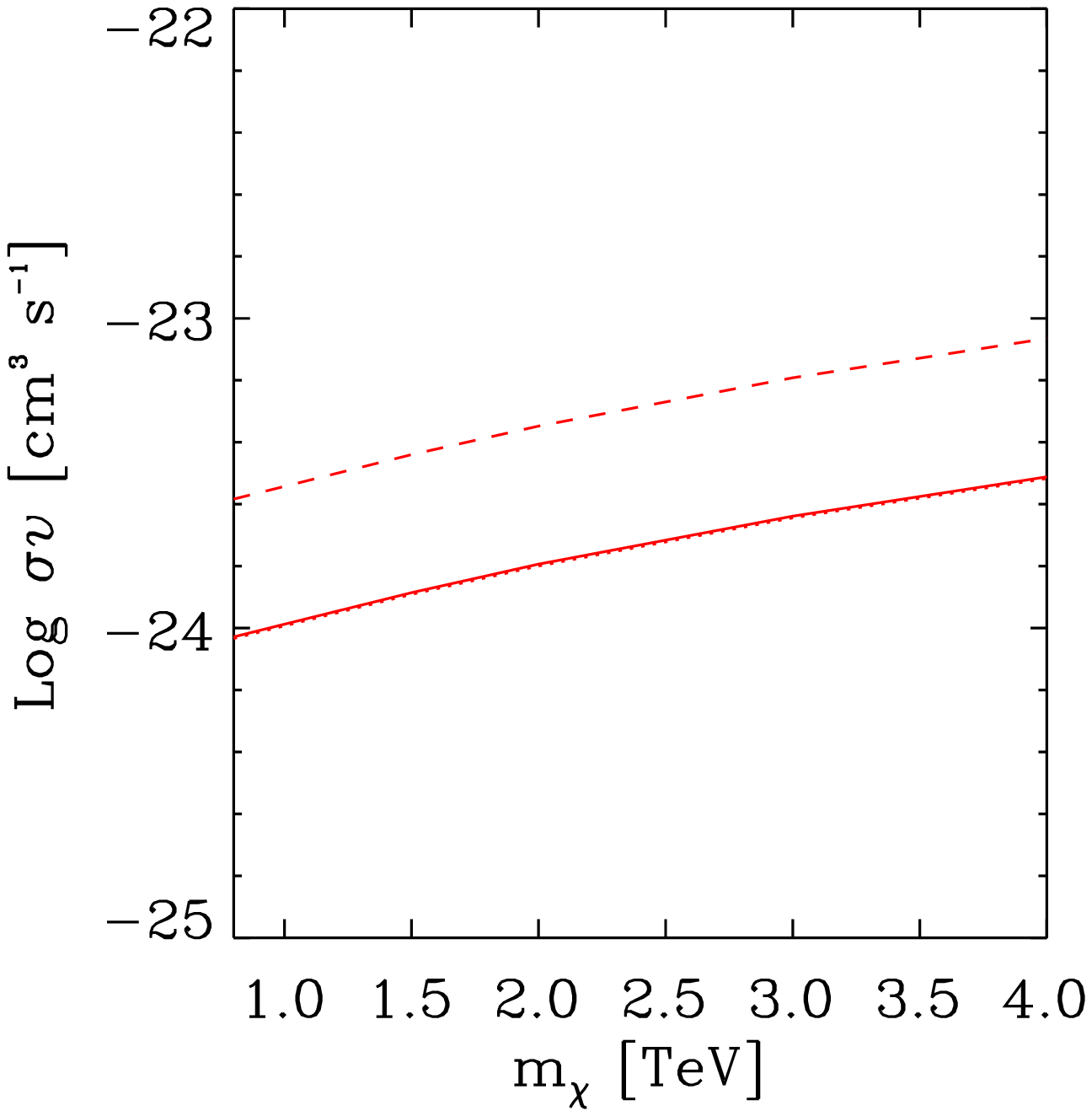}
\caption{Constraints for the Wino ({\it left panel}) and the heavy ({\it right panel}) DM models cross sections as function of the particle mass assuming simulated AMS-02 data for $\bar{p}$. Solid, dotted and dashed lines refers to Einasto, NFW and Burkert DM density profiles respectively.}
\label{fig:ams_constraints}
\end{figure}

We see that even under these pessimistic assumptions the uncertainty on the total ${\bar p}$ should not prevent to detect an excess on the background due to DM annihilation even with small relatively small cross section.  
This comes with the assumption that the secondary production cross section does not cause a break in the background spectrum in the same energies ($\sim$ 100 GeV).
In Fig.~\ref{fig:ams_constraints} (left panel),  we represent the projected AMS-02 sensitivity to Wino-like DM models for the KRA propagation model after 1 years of data taking. 
These plots have been performed following an approach similar to that used to derive the constraints showed in the previous section: for each choice of the DM mass we determined the annihilation cross section such that the total ${\bar p}$ flux exceeds the secondary flux computed for the each of those models  by $3 \sigma$.   
Although these plots do not account for systematical errors, it  is evident by comparing them with those in Fig.~\ref{fig:constraints_wino} that AMS-02 has the potential to improve on the sensitivity on the annihilation  cross sections by more than one order of magnitude. 
In Fig.~\ref{fig:ams_constraints} (right panel) we also show the analogous constraints computed for the heavy dark matter model which annihilate into $\mu^±$ at tree level. From this figure the reader can see as AMS-02 should have the capability to confirm, or to reject, those models as a solution of the PAMELA positron anomaly recently confirmed by Fermi-LAT \cite{Abdo:2009zk}. Yet, we note that we have not taken into account the systematic errors at the highest energies, of proton spill-over and a possible fast increase of $\Delta E/E$ with E (taking it constant instead), since no such information is publicly available. Since these systematic errors can increase the $\bar{p}$ flux, how weaker the DM constraints will actually be will strongly depend on the AMS-02 performance above 100 GeV.   

\subsection{The effect of a break in the CR spectra}

Recent data from the CREAM ballon-borne experiment \cite{Yoon:2011zz} seems to confirm earlier suggestions  (see e.g.~\cite{Panov:2006kf}) that CR nuclei spectra at few TeV/nucleon are harder than in the 10-100 GeV range and that helium and heavier nuclei spectra are harder than the proton spectrum at corresponding energies. As we mentioned in the above, recently PAMELA measured a break in the p and He CR fluxes at rigidity $R \simeq 230$~GV~\cite{Adriani:2011cu} which, if extrapolated to higher energies, is compatible with CREAM results. If confirmed, those features may have relevant implications for the secondary antiproton flux shown also in \cite{Donato:2010vm}, hence for DM indirect constraints.

Here we discus two distinct possibilities for the origin of such a break, and how CR antiprotons measured by AMS-02 could be used as a probe to discriminate among them.

\begin{figure}[tbp]
\centering
\includegraphics[width=0.9\textwidth]{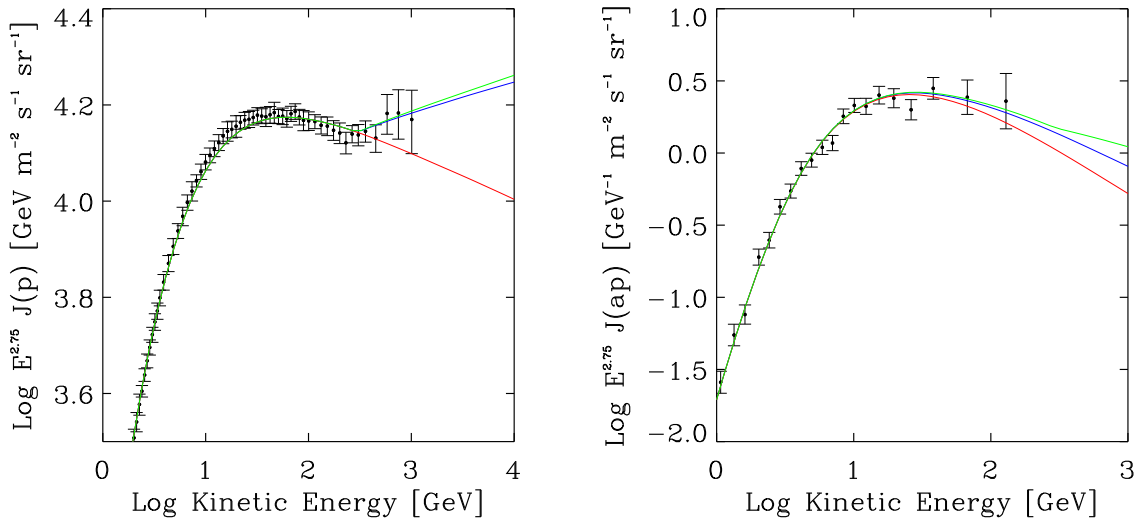}
\caption{{\it Left panel}: The $p$ flux obtained for the KRA model assuming a break in the injection spectra (blue) and a break in the diffusion coefficient (green), compared with the case with no breaks (red). {\it Right panel}: The $\bar{p}$ obtained for the different scenario.}
\label{fig:testbreak}
\end{figure}

The first possible explanation for the break at rigidity $R\simeq 230$~GV, shown also in Fig.~\ref{fig:testbreak} (left panel) for the p flux is that this break comes from the injection of CR $p$ and heavier nuclei into the ISM. That could be because at the acceleration sites the formation of a precursor may take place as has been suggested semianalytic work on the diffusive shock acceleration~\cite{2002APh....16..429B, Blasi:2006wj, Caprioli:2010ne, Caprioli:2010uj}. The presence of a precursor would lead in higher energy particles being compressed more (on average) and thus accelerated to a harder spectrum than the lower energy particles \cite{2002APh....16..429B, Caprioli:2010uj}. Alternatively a second population of SNRs could be emerging at $\sim$ 230 GV. Both cases are modeled as ``a break in the injection spectra'', shown as blue lines in Fig.~\ref{fig:testbreak}. The second possible explanation is that at $\sim$ 230 GV we observe a change in the turbulence power spectrum of the ISM as seen by CRs, which could be Kraichnan type at low $R$ and Kolmogorov type at high $R$, which we show as green lines in Fig.~\ref{fig:testbreak}. 

Both cases can explain as we show in Fig.~\ref{fig:testbreak} (left panel) the observed break at the protons (shown) and He (not shown). As antiprotons are produced by the CR $p$, He and metals, the observed break at 230 GV, should lead in a first soft break (hardening of the spectrum), shown in Fig.~\ref{fig:testbreak} (left panel), compared to the $\bar{p}$ flux predicted if $p$ and He did not have any break at 230 GV (red line). That first soft break is the case for both the break in the injection and the break in the diffusion coefficient(Fig.~\ref{fig:testbreak}). Yet in the case of the break in the diffusion coefficient, we should also observe from the propagation of the secondary $\bar{p}$s in the ISM a second hardening at $\simeq$230 GV, shown in Fig.~\ref{fig:testbreak}(right panel). On the basis of this argument, in \cite{Donato:2010vm} it was noticed that if AMS-02 observes an hardening in the $\bar{p}$ spectrum at the same rigidity where PAMELA has observed the break in $p$ and He it would confirm the break in the diffusion coefficient.

A complementary analysis has also been carried out in \cite{Cholis:2011un} where the authors concluded that the differences between the break in the injection versus the break in the diffusion are too small in the diffuse galactic flux at middle and high latitudes and $E_{\gamma} > 100$ GeV, to give a conclusive result for one vs the other scenario. Thus $\bar{p}$s may actually be the best probe to understand the origin of the 230 GV break.

\section[]{Conclusions} \label{sec:concl}

In this article, we have explored various sources of uncertainty which affect the search of the antiproton cosmic ray component possibly produced by dark matter annihilations in the Galaxy. We mainly focused on astrophysical uncertainties, due to CR propagation and DM spatial distribution.    

For this purpose, we have consistently computed the local flux of antiprotons produced by DM and by interactions of p/He nuclei with the ISM for several numerical diffusion models.  We fixed model parameters under the prescription that they provide good statistical fits ($\chi_{\rm red}^2 < 1$) of recently updated  B/C and proton data under different conditions which may affect the secondary and especially the DM ${\bar p}$ fluxes. In particular, we have studied different rigidity dependencies of the diffusion coefficient, a wide range of galactic halo thickness and also considered a model with a strong convection. All these models are in very good agreement with the new PAMELA data \cite{Adriani:2010rc} and they also give a rather good description of the $^4$He spectrum measured by the same experiment above few GeV. 

For what concerns secondary antiprotons, we found a rather weak variance (less than $30\%$ between 0.1 and 100 GeV) of their flux on the choice of the propagation model. 
This is explained by the fact that their production mechanism and energy losses are similar to those of secondary nuclei on which the models are tuned. 
The agreement between the predicted secondary ${\bar p}$ and the experimental data (especially those of PAMELA~\cite{Adriani:2010rc}) are very good with the only marginal exception of the KOL model (see also \cite{Evoli:2008dv}).  
Therefore, in agreement with previous results, we conclude that from the presently available antiproton data it comes no reason to invoke a new contribution to the flux and that a possible DM contribution to the antiproton flux must be subdominant.  

Then we compared numerical predictions with experimental data to get constrains on DM annihilation models for different propagation models and DM density profiles. 
We considered three classes of DM models discussed in Sec.~\ref{sec:dm} recently investigated in connection to hints of DM signals in other detection channels but potentially giving a sizable antiproton flux as well.  For each of those models we considered three DM profiles (Einasto, Navarro-Frenk-White and Burkert) which are often adopted in the related literature. 
We found a strong dependence, up to a factor of about 50, of the DM ${\bar p}$ flux on the choice of the propagation model. This variance is dominated by the large uncertainty on the propagation parameters, most importantly by that on the diffusion halo scale height and it is even larger than the uncertainty due to the choice of the DM profile. 

In order to interpret these results, we have investigated how the relative antiproton flux changes as a function of the source position in the Galaxy for different propagation models. In qualitative agreement with findings by other groups employing semianalytic techniques to solve the transport equations,\cite{Taillet:2002ub, Maurin:2002uc} we found that secondary nuclei and antiprotons reaching the Earth are mainly produced within a distance of few kpc from the Earth position. They are therefore very weakly affected by possible variations of the propagation conditions in the inner Galaxy where most of the DM antiprotons are produced.  In order to test the possible consequences of this issue on the DM constraints, we exploited the features of the \dragon\ tool that allowed us to consider also some nonstandard, though physically motivated, diffusion models in which the physical conditions in the inner 3 kpc of the Galaxy are different from the rest of the disk. We found a significant effect on the DM contribution without affecting the local observables like B/C and protons that are mostly used to determine the propagation model parameters. Other than being by itself another source of uncertainty, this also means that the propagation uncertainty in DM antiproton predictions cannot be reduced beyond the one given by nonstandard propagation models even if new high precision measurements of the local nuclear observables will be available in the future. Only a comprehensive study including local and nonlocal (e.g.,~$\gamma$-ray) observables may succeed in reducing safely the propagation uncertainties.

With these limitations in mind, we derived new constraints on DM annihilation cross section for a set of propagation models adopting radially uniform propagation properties. 
In spite of the large astrophysical uncertainties discussed above, our constraints already exclude some models which rose a wide interest in the recent literature, such as $\approx$ 200 GeV Wino models~\cite{Kane:2009if} suggested in connection to the rise of the positron fraction measured by PAMELA, and light binolike DM models in connection to the CoGent and DAMA signals. It is worth reminding here that our analysis accounts for electroweak corrections to DM annihilation spectra, which significantly affect the ${\bar p}$ flux produced by the annihilation of heavy ($m_\chi \gg 100$ GeV) DM particles \cite{Ciafaloni:2010ti}. These corrections are especially relevant for the very heavy WIMP scenario, which was proposed as a possible interpretation of the PAMELA positron anomaly ~\cite{Adriani:2008zr}, as they make possible an independent test of ``{\it leptophilic}" models in the ${\bar p}$ channel. 

Beside considering the astrophysical uncertainties mentioned above, we also explored the effects of the uncertainties on the gas distribution, the spallation cross sections and the local distribution of dark matter where we also studied the effects of a dark disk component. 
Upon confirming previous studies, our results show that these uncertainties are relatively less important with respect to the variance obtained by adopting different propagation models.

We also compared our numerical results with semianalytical solutions widely used in the related literature. 
In most cases we found relatively small, though not negligible, discrepancies (up to $25~\%$ or larger). 

At the end of this paper we estimated the projected sensitivity of the AMS-02 space observatory to some of the DM models considered in the above.  
We showed as the interplay of its accurate CR nuclei and antiproton measurements should be able to improve dramatically the sensitivity to DM models with respect to the constraints derived in this work.   
 
Furthermore, we have discussed the implications from the recently found rigidity break in the protons and He CR spectra~\cite{Adriani:2011cu}. We have addressed the possibility of discriminating whether the break is in the injection spectrum (connected to either acceleration effects in the sources, or to the presence of an extra population of primary sources injecting CRs with harder spectra) or in the energy dependence of the diffusion coefficient by using forthcoming observation up to $\sim$500 GeV energy range with smaller statistical errors.

\section*{Acknowledgments}
The authors would like to thank Mirko Boezio, Christoph Weniger and Pasquale D. Serpico for the interesting discussions we shared. 
LM and CE acknowledge support from the State of Hamburg, through the Collaborative Research program ``Connecting Particles with the Cosmos''.
CE warmly thanks DESY, Hamburg University and GGI for kind hospitality during the preparation of this work. DG thanks the CERN theory group for kind hospitality and financial support during part of the preparation of this work. 

\bibliography{exoticap}
\end{document}